\documentclass{emulateapj}

\usepackage{multirow}
\usepackage{graphics}
\usepackage{amsmath, amsthm, amssymb}

\usepackage{epsfig}
\tightenlines
\newcommand \lsim{\mathrel{\rlap{\lower4pt\hbox{\hskip1pt$\sim$}}
    \raise1pt\hbox{$<$}}}
\newcommand \gsim{\mathrel{\rlap{\lower4pt\hbox{\hskip1pt$\sim$}}
    \raise1pt\hbox{$>$}}}

\newcommand     \kpc    {\,{\rm kpc}}

\newcommand{\sm}{{M_\odot}}

\newcommand{\beq}{\begin{equation}}
\newcommand{\eeq}{\end{equation}}
\newcommand{\beqa}{\begin{eqnarray}}
\newcommand{\eeqa}{\end{eqnarray}}

\newlength{\figwidth}
\countdef\decade=200
\advance\decade by \year
\countdef\hours=201
\advance\hours by \time
\divide\hours by 60
\countdef\mins=202
\advance\mins by \hours
\multiply\mins by 60
\multiply\hours by 100
\countdef\miltime=203
\advance\miltime by \hours
\advance\miltime by \time
\advance\miltime by -\mins

\begin{document}

\title{Mid-Infrared Extinction Mapping of Infrared Dark Clouds:\\ Probing the Initial Conditions for Massive Stars and Star Clusters}

%\centerline{DRAFT: \today}

\author{Michael J. Butler}
%\affil{Dept. of Astronomy, University of Florida, Gainesville, FL 32611, USA}
\author{Jonathan C. Tan}
%\affil{Department of Astronomy, University of Florida, Gainesville, FL 32611, USA; butler85@astro.ufl.edu, jt@astro.ufl.edu}
%emulate
\affiliation{Department of Astronomy, University of Florida, Gainesville, FL 32611, USA;\\ butler85@astro.ufl.edu, jt@astro.ufl.edu}

\begin{abstract}
Infrared Dark Clouds (IRDCs) are cold, dense regions of giant
molecular clouds that are opaque at wavelengths $\sim 10\:{\rm \mu m}$
or more and thus appear dark against the diffuse Galactic background
emission. They are thought to be the progenitors of massive stars and
star clusters.  We use 8~$\rm \mu m$ imaging data from {\it Spitzer}
GLIMPSE to make extinction maps of 10 IRDCs, selected to be relatively
nearby and massive. The extinction mapping technique requires
construction of a model of the Galactic IR background intensity behind
the cloud, which is achieved by correcting for foreground emission and
then interpolating from the surrounding regions. The correction for
foreground emission can be quite large, up to $\sim 50\%$ for clouds
at $\sim 5\:{\rm kpc}$ distance, thus restricting the utility of this
technique to relatively nearby clouds. We investigate three methods
for the interpolation, finding systematic differences at about the
10\% level, which, for fiducial dust models, corresponds to a mass
surface density $\Sigma = 0.013\:{\rm g\:cm^{-2}}$, above which we
conclude this extinction mapping technique attains validity. We
examine the probability distribution function of $\Sigma$ in
IRDCs. From a qualitative comparison with numerical simulations of
astrophysical turbulence, many clouds appear to have relatively narrow
distributions suggesting relatively low ($<5$) Mach numbers and/or
dynamically strong magnetic fields. Given cloud kinematic distances,
we derive cloud masses. Rathborne, Jackson \& Simon identified cores
within the clouds and measured their masses via mm dust emission. For
43 cores, we compare these mass estimates with those derived from our
extinction mapping, finding good agreement: typically factors of
$\lesssim 2$ difference for individual cores and an average systematic
offset of $\lesssim 10\%$ for the adopted fiducial assumptions of each
method. We find tentative evidence for a systematic variation of these
mass ratios as a function of core density, which is consistent with
models of ice mantle formation on dust grains and subsequent grain
growth by coagulation, and/or with a temperature decrease in the
densest cores.
\end{abstract}


\section{Introduction}\label{S:intro}

Large fractions of stars form in clusters from the densest {\it
clumps} within giant molecular clouds (GMCs) (Lada \& Lada
2003). These regions are also responsible for the birth of essentially
all massive stars (de Wit et al. 2005). It is possible that our own
solar system formed in such a region near a massive star (Hester et
al. 2004), so the process of massive star and star cluster formation
may be directly involved in our own origins. Understanding the
formation of star clusters is also important as a foundation for
understanding global galactic star formation rates (Kennicutt 1998)
and thus the evolution of galaxies.

In spite of this importance, there are many gaps in our knowledge of
how massive stars and star clusters are formed. For massive star
formation there is a basic debate about whether the process is simply
a scaled-up version of low-mass star formation from gas cores (Shu,
Adams, Lizano 1987), albeit requiring the high pressures found in the
centers of star-forming clumps (McKee \& Tan 2003), or whether a
qualitatively different mechanism is involved, such as stellar
collisions (Bonnell, Bate, Zinnecker 1998) or competitive Bondi-Hoyle
accretion (Bonnell et al. 2001; Bonnell, Vine, \& Bate 2004). A
prediction of the scenario involving formation from cores is the
presence of massive, gravitationally-bound starless cores as initial
conditions. 

For star cluster formation there is a debate about whether the {\it
protocluster} (or {\it star-forming clump}) is in quasi-virial
equilibrium (Tan, Krumholz, \& McKee 2006) or is undergoing rapid
global collapse (Elmegreen 2000, 2007; Hartmann \& Burkert 2007). This
corresponds to a debate about the timescale of star cluster formation:
does it take many or just a few free-fall times?

To help resolve these issues we require knowledge about the initial
conditions of massive star and star cluster formation. The
star-forming clumps that are the sites of these processes have
typically been identified from the radiative emission and activity of
their young stars: e.g. radio emission from ultracompact \ion{H}{2}
regions created by massive stars or maser emission from hot, dense gas
near massive protostars. Unfortunately by the time this activity
signposts the region, it has typically also erased much of the memory
of the initial conditions from the system. This is especially true of
protostellar outflows, which have a mechanical power large enough to
significantly stir the gas of the star-forming clump and halt any
large scale gravitational collapse (Nakamura \& Li 2007).

The initial conditions for massive star and star cluster formation are
expected to be dense, cold cores and clumps of gas. In some formation
models there may be of order hundreds of solar masses of gas
compressed to within a few tenths of a parsec. For a spherical cloud
the mean mass surface density, $\Sigma$, in this case is 
\beq 
\label{Sigma}
\Sigma \equiv \frac{M_c}{\pi R_c^2} = 0.665
\frac{M_c}{100\sm}\left(\frac{R_c}{0.1{\rm pc}}\right)^{-2}\:{\rm
g\:cm^{-2}}.  
\eeq 
For reference, $\Sigma=1\:{\rm g\:cm^{-2}}$ corresponds to
$4800\:M_\odot\:{\rm pc^{-2}}$, $N_{\rm H}=4.3\times 10^{23}\:{\rm
cm^{-2}}$ and $A_V\simeq 230$~mag, for local diffuse ISM dust
properties (e.g. Draine 2003). The extension of the extinction law
into the mid-infrared (MIR) is somewhat controversial and uncertain
(Lutz et al. 1996; Draine 2003; Indebetouw et al. 2005; Rom\'an-Z\'u\~niga
et al. 2007), but nevertheless, such column
densities are expected to correspond to several magnitudes of
extinction at 8~$\rm \mu m$.

Such Infrared Dark Clouds (IRDCs) have been identified in images of
the Galactic plane from the {\it Infrared Space Observatory} (ISO)
(P\'erault et al. 1996), the {\it Midcourse Space Experiment} (MSX)
(Egan et al. 1998). Simon et al. (2006a) identified about 10,000
potential IRDCs from intensity contrast features in the MSX
survey. Simon et al. (2006b) identified the $^{13}$CO emission from
about 300 of the darkest of these in the Galactic Ring Survey (GRS)
(Jackson et al. 2006), thus deriving their kinematic
distances. Rathborne, Jackson, \& Simon (2006) surveyed the 1.2~mm
dust continuum emission from 38 of these IRDCs, identifying 140
mm-emission cores within these clouds.

The temperatures in IRDCs are measured to be $\lesssim 20$~K (Carey et
al. 1998). High deuteration fractions have been reported by Pillai et
al. (2007). Under these conditions one expects high depletion of
volatiles onto ice mantles of dust grains (Dalgarno \& Lepp 1984).
Larger molecular line surveys of IRDCs have been carried out by, for
example, Ragan et al. (2006) and Sakai et al. (2008).

In this paper we present a method of extinction mapping of IRDCs using
the diffuse Galactic IR emission observed by {\it Spitzer} as the background
source. This method complements that based on measuring the extinction
to individual stars (e.g. Rom\'an-Z\'u\~niga et al. 2007), by
providing a measurement of the extinction at the location of almost
every pixel in the cloud image, including at very high column
densities, thus allowing a detailed investigation of the cloud structure.

\section{IRDC Sample Selection}\label{S:sample}

%\begin{deluxetable}{lcccccccccc}
%emulate
\begin{deluxetable*}{lcccccccccc}
\tabletypesize{\footnotesize}
\tablecolumns{10}
\tablewidth{0pt}
\tablecaption{Infrared Dark Cloud Sample\tablenotemark{a}}
\tablehead{\colhead{Cloud Name}                                              &
  \colhead{$l$}  &
  \colhead{$b$}  &
           \colhead{Distance}                                                   &
	   \colhead{$R_{\rm eff}$}  &
	   \colhead{$e$}  &
	   \colhead{P.A.}  &
           \colhead{$f_{\rm fore}$}                                                              &
           \colhead{$\bar{\Sigma}_{\rm SMF}$\tablenotemark{b}}                                  &
           \colhead{$M_{\rm LMF}$}                                  &
           \colhead{$M_{\rm SMF}$}                                  \\
           \colhead{}                                                                      &
           \colhead{($^{\circ}$)}                                                                   &
           \colhead{($^{\circ}$)}                                                                   &
           \colhead{$(\kpc)$}                                                                &
           \colhead{(pc)}                                                                &
           \colhead{}                                                                      &
           \colhead{($^{\circ}$)}                                                                   &
           \colhead{}                                                                &
           \colhead{$\rm (g\:cm^{-2})$}                                                                &
           \colhead{$(\sm)$}                                                                 &
           \colhead{$(\sm)$}                                                                 }
\startdata
%\sidehead{Cloud A}
A (G018.82$-$00.28) & 18.822 & $-$0.285  & 4.8 & 10.4 & 0.961 & 74 & 0.209 & 0.0355 & 6,700 & 7,600                   \\
%\sidehead{Cloud B}
B (G019.27$+$00.07) & 19.271 & 0.074     & 2.4 & 2.71 & 0.977 & 88 & 0.075 & 0.0387 & 930 & 830                   \\
%\sidehead{Cloud C}
C (G028.37$+$00.07) & 28.373 & 0.076     & 5.0 & 15.4 & 0.632 & 78 & 0.266 & 0.0527 & 27,000 & 42,000                   \\
%\sidehead{Cloud D}
D (G028.53$-$00.25) & 28.531 & $-$0.251  & 5.7 & 16.9 & 0.968 & 60 & 0.327 & 0.0418 & 17,400 & 27,000                   \\
%\sidehead{Cloud E}
E (G028.67$+$00.13) & 28.677 & 0.132  & 5.1 & 11.5 & 0.960 & 103 & 0.276 & 0.0543 & 15,100 & 19,400                   \\
%\sidehead{Cloud F}
F (G034.43$+$00.24) & 34.437 & 0.245     & 3.7 & 3.50 & 0.926 & 79 & 0.193 & 0.0371 & 1,770 & 1,670                   \\
%\sidehead{Cloud G}
G (G034.77$-$00.55) & 34.771 & $-$0.557  & 2.9 & 3.06 & 0.953 & 95 & 0.140 & 0.0420 & 1,050 & 1,140                   \\
%\sidehead{Cloud H}
H (G035.39$-$00.33) & 35.395 & $-$0.336  & 2.9 & 9.69 & 0.951 & 59 & 0.142 & 0.0262 & 4,000 & 6,800                   \\
%\sidehead{Cloud I}
I (G038.95$-$00.47) & 38.952 & $-$0.475  & 2.7 & 3.73 & 0.917 & 64 & 0.141 & 0.0616 & 880 & 1,490                   \\
%\sidehead{Cloud J}
J (G053.11$+$00.05) & 53.116 & 0.054     & 1.8 & 0.755 & 0.583 & 50 & 0.121 & 0.0699 & 108 & 80                   \\

\enddata

\tablenotetext{a}{Coordinate names, Galactic coordinates, kinematic
distances, effective radii (of equal area circles), eccentricities and
position angles of fitted ellipses are from Simon et al. (2006a).}  

\tablenotetext{b}{This estimate of mean mass surface density, used to
normalize the distributions in Fig.~\ref{fig:sigdistrnormnew}, is the
areal average of those pixels for which values of $\Sigma_{\rm SMF}>0$
are derived. Estimates of a mean mass surface density based on $M_{\rm
SMF}$ and $R_{\rm eff}$ are typically much smaller because of the
regions inside the clouds ellipse with $\Sigma_{\rm SMF}\leq 0$.}
\label{tb:clouds}
%\end{deluxetable}
%emulate
\end{deluxetable*}

Considering {\it Spitzer} Galactic Legacy Mid-Plane Survey
Extraordinaire (GLIMPSE) (Benjamin et al. 2003) 8~$\rm \mu m$
(i.e. Infrared Array Camera [IRAC] band 4) images, we chose 10 IRDCs
from the sample of Rathborne et al. (2006), selecting those that were
relatively nearby, massive, dark (i.e. showing high contrast compared
to the surrounding diffuse emission), and/or with relatively simple
surrounding diffuse emission.  The properties of these clouds are
listed in Table \ref{tb:clouds}. Apart from being relatively nearby,
this sub-sample is in fact fairly representative of the full 38 cloud
sample of Rathborne et al. (2006).

Simon et al. (2006a) fit ellipses to each cloud based on MSX
images. While these ellipses are often not particularly accurate
descriptions of the IRDC shapes, we will utilize them as convenient
measures of the approximate sizes and shapes of the clouds, especially
for the small scale median filter method of estimating the background
radiation (\S\ref{S:m2}).

The {\it Spitzer} telescope has an angular resolution (PSF FWHM) of
about 2\arcsec\ at $\rm 8\:\mu m$, which corresponds to a linear scale
of 0.029~pc for a cloud at a distance of 3~kpc. Note, GLIMPSE images
are processed to a pixel scale corresponding to an angular resolution
of 1.2\arcsec.

%When choosing candidate IRDCs for our analysis, criteria such as
%distance, size, and complexity of the surrounding region were taken
%into consideration.  The ideal IRDC would be relatively close(within a
%few $\kpc$), large, and in a simple region free of contamination from
%bright extended sources or other dark material not belonging to the
%cloud.  It should also have a high level of contrast compared to the
%background, such that its boundaries are well-defined.  From the
%sample of Rathborne et al.(2006), the six IRDCs that were closest to
%fitting these criteria were chosen.  These clouds are listed in Table
%1.

\section{IRDC Extinction Mapping Methods}\label{S:methods}

\begin{figure*}
\begin{center}
\includegraphics[width=6in]{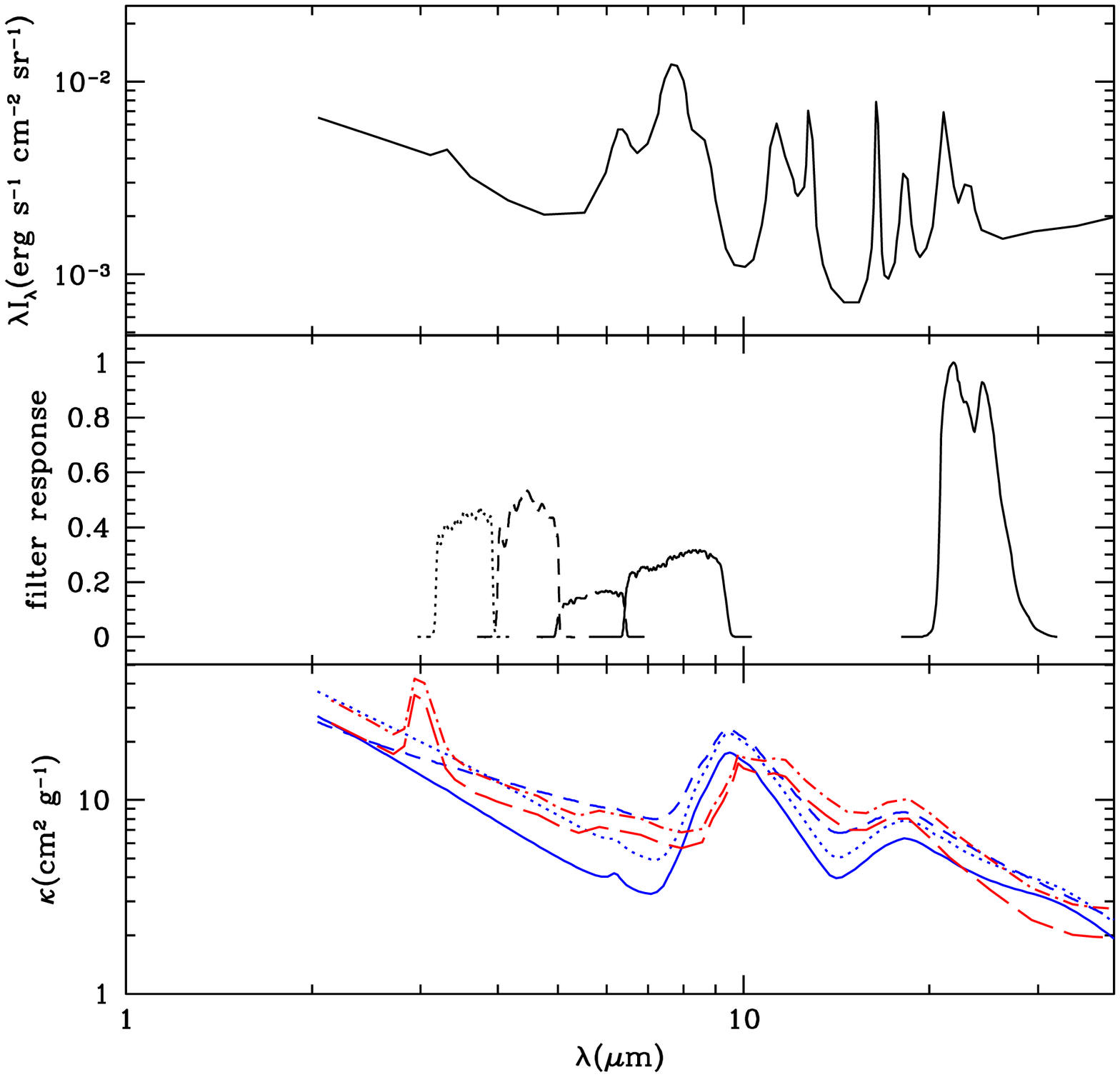}%irext.eps 
\caption{{\it Top:} Adopted spectrum of diffuse IR background from the
``MIRS'' region in the Galactic plane at $l\simeq44^\circ$ modeled by
Li \& Draine (2001). {\it Middle:} Filter response of IRAC bands 1-4
and the MIPS 24~$\rm \mu$m band. {\it Bottom:} Dust opacities (per gas
mass) from Weingartner \& Draine (2001) as updated by Draine (2003):
Model with $R_V=3.1$ ({\it solid line}), $R_V=5.5$ ({\it dotted
line}), $R_V=5.5$ case B ({\it dashed line}). These opacities include
scattering, which is about a 40\% of total effect at 3~$\rm \mu$m, but
only a 2\% effect at 8~$\rm \mu$m. The {\it long-dashed} line shows
the thin ice mantle MRN model (only contribution from $\kappa_{\rm
abs}$ per gas mass is shown) of Ossenkopf \& Henning (1994). The {\it
dot-dashed line} shows this model after $10^5$~yr of coagulation at a
density of $n_{\rm H}=10^6\:{\rm cm^{-3}}$. The background and
filter-weighted opacities are listed in Table~\ref{tb:kappa}.
\label{fig:irext}}
\end{center}
\end{figure*}

%\begin{deluxetable}{lccccc}
%emulate
%\rotate
\begin{deluxetable*}{lccccc}
\tabletypesize{\footnotesize}
\tablecolumns{6}
\tablewidth{0pt}
\tablecaption{Spitzer Telescope Band and Background-Weighted Dust Opacities Per Gas Mass ($\rm cm^{2}\:g^{-1}$)}
\tablehead{\colhead{Dust Model\tablenotemark{a}}                                              &
           \colhead{IRAC Band 1} &  
           \colhead{IRAC Band 2} &  
           \colhead{IRAC Band 3} &  
           \colhead{IRAC Band 4} &  
           \colhead{MIPS Band 1} \\
	   \colhead{} &
	   \colhead{$\rm 3.5 \mu m$\tablenotemark{b}} &
	   \colhead{$\rm 4.5 \mu m$} &
	   \colhead{$\rm 5.9 \mu m$} &
	   \colhead{$\rm 7.8 \mu m$} &
	   \colhead{$\rm 23.0 \mu m$}
}
\startdata

WD01 $R_V=3.1$ & 10.02 & 6.26 & 4.16 & 6.25 & 4.50\\
WD01 $R_V=3.1$ flat IR bkg\tablenotemark{c} & 9.78 & 6.20 & 4.25 & 7.71 & 4.24\\
WD01 $R_V=5.5$ & 15.56 & 10.24 & 6.55 & 8.27 & 5.54 \\
WD01 $R_V=5.5$ case B & 14.23 & 11.49 & 9.25 & 10.96 & 6.01\\
OH94 thin mantle, 0~yr & 17.52 (12.69) & 10.68 (8.69) & 7.78 (7.06) & 6.26 (6.15) & 4.43\\
OH94 thin mantle, $10^5$yr, $10^6{\rm cm^{-3}}$ & 22.41 (16.23) & 13.31 (10.83) & 9.44 (8.56) & 7.48 (7.34) & 6.23\\
\enddata
\tablenotetext{a}{References: WD01 - Weingartner \& Draine (2001); OH94 - Ossenkopf \& Henning (1994), opacities have been scaled from values in parentheses to include contribution from scattering.}
\tablenotetext{b}{Mean wavelengths weighted by filter response and background spectrum.}
\tablenotetext{c}{No weighting made for spectrum of Galactic diffuse emission.}
\label{tb:kappa}
%\end{deluxetable}
%emulate
\end{deluxetable*}

The extinction mapping technique requires knowing the intensity of
radiation directed towards the observer at a location behind the cloud
of interest, $I_{\nu,0}$, and just in front of the cloud,
$I_{\nu,1}$. Then for negligible emission in the cloud and a
simplified 1D geometry,
\begin{equation}
\label{eq:radtrans}
I_{\nu,1}=I_{\nu,0} \:{\rm exp}(-\tau_\nu),
\end{equation}
where the optical depth $\tau_\nu =\kappa_\nu \Sigma$, where $\kappa_\nu$
is the total opacity at frequency $\nu$ per unit gas mass and $\Sigma$ is the gas mass
surface density.

To evaluate $\kappa_\nu$ appropriate for the various intensities
measured by the {\it Spitzer Space Telescope}, namely the 4 IRAC bands
and the Multiband Imaging Photometer (MIPS)~24~$\rm \mu$m band, we
adopt a spectrum of the diffuse Galactic IR emission from Li \& Draine
(2001), the filter response functions of the IRAC and MIPS bands, and
perform an intensity and filter response weighted average of the
opacity of various dust models (Figure~\ref{fig:irext} \& Table
\ref{tb:kappa}). Uncertainties in the dust models include the extent
to which ice mantles have formed on the grains and the extent to which
the grains have undergone coagulation.

%The optical depth can then be
%converted to column density, $\sum$ ($g$/$cm^2$), by dividing by the dust
%opacity, $\kappa$(Weingartner and Draine 2001).  We use a $\kappa$ value of
%728.129. Since this value of $\kappa$ is only for dust, $\sum$ must be
%multiplied by a factor of 123.6 in order to obtain the total surface
%density(\ref{fig:method1}d).

%IRDCs are cold ($T\lesssim20$~K) relatively dense ($n_{\rm H}\sim
%10^5\:{\rm cm^{-3}}$) structures, and this environment may affect the
%properties of dust grains, especially the formation of ice mantles and
%enhanced coagulation, leading to larger, more porous grains
%(e.g. Ossenkopf \& Henning 1994).

The IR background in the wavelength range probed by the IRAC bands
receives its greatest contribution from the diffuse ISM (transiently
heated small grains) at Band 4, i.e $\sim 8\:{\rm \mu m}$, compared to
that from background stars. Individual stellar sources become much
more prominent in the GLIMPSE images at the shorter wavelengths. Thus
in this paper we restrict our analysis to Band 4 images, leaving
analysis at other wavelengths and the wavelength dependence of
extinction for a future study. 

In the $\rm 8\: \mu m$ band, we estimate a range of dust opacities per
gas mass of $6.3-11\:{\rm cm^2\:g^{-1}}$, and adopt $\kappa_\nu=7.5\:{\rm
cm^2\:g^{-1}}$, which is formally closest to the model of Ossenkopf \&
Henning (1994) with thin ice mantles that have undergone coagulation
for $10^5\:{\rm yr}$ at a density of $n_{\rm H}=10^6\:{\rm cm^{-3}}$
(or approximately equivalent to $10^6$~yr at $n_{\rm H}=10^5\:{\rm
cm^{-3}}$, etc.).

We estimate $I_{\nu,0}$ via interpolation from the regions surrounding
the particular IRDC of interest. These interpolation methods, which
necessarily involve an averaging over small scale spatial variations
in $I_{\nu,0}$, are described below in \S\ref{S:interpolation}. We
evaluate $I_{\nu,1}$ from the observed intensities derived from the
cloud images. First we consider the effects of foreground dust emission.

\subsection{Correction for Foreground Dust Emission}

Our determinations of both $I_{\nu,0}$ and $I_{\nu,1}$ are potentially
affected by foreground emission from hot dust. We estimate the size of
this effect given the (kinematic) distance to the cloud and a model
for the Galactic distribution of hot dust emission, assuming it is the
same as the distribution of the Galactic surface density of OB
associations (McKee \& Williams 1997):
\begin{equation}
\Sigma_{\rm OB} \propto {\rm exp} \left(-\frac{R}{H_{\rm R}}\right),
\end{equation}  
where $R$ is the galactocentric radius and $H_R=3.5$~kpc is the radial
scale length. For each IRDC, given its distance and Galactic
longitude, we calculate the ratio of the column of hot dust between
the solar position (at $R=8$~kpc) and the total column extending out to
a galactocentric radius of 16~kpc. This ``Foreground Intensity
Ratio'', $f_{\rm fore}$, is listed in Table~\ref{tb:clouds} for each
IRDC. Then we derive an estimate of the true intensity of the
radiation field just behind the cloud, $I_{\nu,0}$, from that measured
via interpolation of the cloud images, $I_{\rm \nu,0,obs}$, via
\beq
I_{\nu,0} = (1-f_{\rm fore}) I_{\rm \nu,0,obs}.
\eeq
We also estimate the true intensity of the radiation field just in
front of the cloud, $I_{\nu,1}$ from that measured directly from the
cloud images, $I_{\rm \nu,1,obs}$, via
\beq
I_{\nu,1} = I_{\rm \nu,1,obs} - f_{\rm fore} I_{\rm \nu,0,obs}.
\eeq

Typical values of $f_{\rm fore}$ are about 15\%, with values up to
33\% for the most distant cloud, D, at 5.7~kpc. As an example of the
size of the corrections to $\Sigma$ resulting from the foreground
subtraction, consider cloud C, for which $f_{\rm fore}=0.266$, $I_{\rm
\nu,0,obs}\simeq 100\:{\rm MJy\:sr^{-1}}$ and $I_{\rm \nu,1,obs}\simeq
50 \:{\rm MJy\:sr^{-1}}$ in the darkest part of the cloud. For these
regions, one would derive $\Sigma \simeq 0.092\:{\rm g\:cm^{-2}}$
without the foreground correction and $\Sigma \simeq 0.152\:{\rm
g\:cm^{-2}}$ with this correction.

Our estimate of $f_{\rm fore}$ is uncertain due to small scale spatial
variations in the hot dust emission in the Galaxy. Also, we are
typically measuring $I_{\rm \nu,0,obs}$ from regions relatively close
to the IRDC of interest. Such regions are likely to overlap with the
GMC hosting the IRDC, and thus have higher than average extinction of
the integrated Galactic background emission. This will cause us to
tend to underestimate $f_{\rm fore}$ and thus $\Sigma$. An upper limit
to $f_{\rm fore}$ is provided by the minimum value of $I_{\rm
\nu,1,obs}$ for each cloud. For example, for Cloud C this is about
$40\:{\rm MJy\:sr^{-1}}$ intensity units, so that for a background of
$100 \:{\rm MJy\:sr^{-1}}$, the maximum value of $f_{\rm fore}\simeq
0.4$. Uncertainties in $f_{\rm fore}$ are one of the major reasons for
restricting the extinction mapping analysis to relatively nearby
clouds.

\subsection{Background Estimation}\label{S:interpolation}

\begin{figure*}
%newcloudC.eps
%newratio1over2.eps
%newcloudCmethod1background.eps
%newcloudCmethod2background.eps
%newmethod1cloudCsigma.eps
%newm2cloudCsigma.eps

\begin{center}
\includegraphics[width=5in]{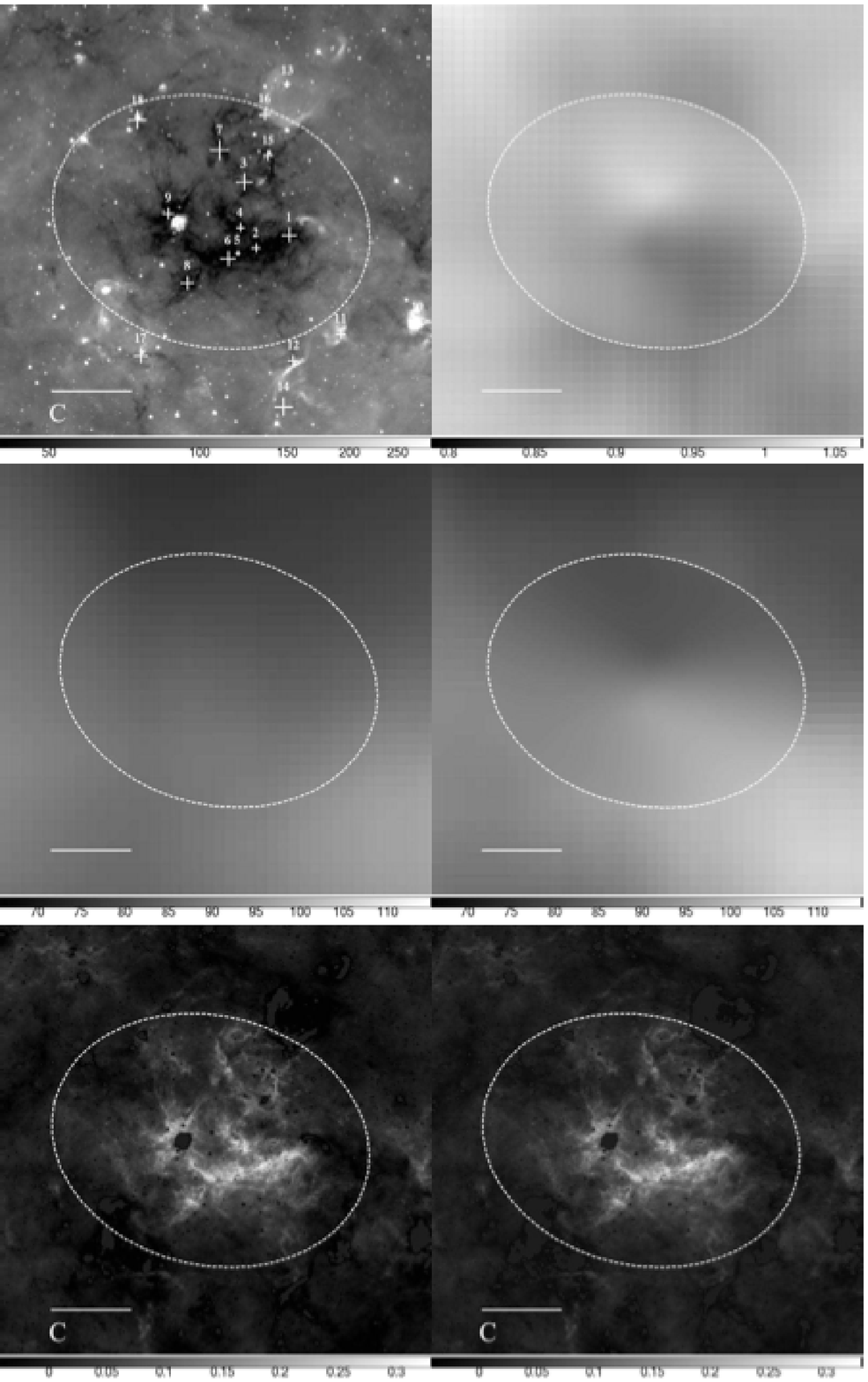}
%$
%\begin{array}{cc}
%\includegraphics[width=2.6in]{f2a_s.eps}
%\includegraphics[width=2.6in]{f2b_s.eps}
%\\
%\includegraphics[width=2.6in]{f2c_s.eps}
%\includegraphics[width=2.6in]{f2d_s.eps}
%\\
%\includegraphics[width=2.6in]{f2e_s.eps}
%\includegraphics[width=2.6in]{f2f_s.eps}
%\end{array}$
\caption{\footnotesize {\it Top left}: Spitzer IRAC 8$\mu m$ image of Cloud C
(G028.37$+$00.07) (logarithmic intensity scale in units of ${\rm
MJy\:sr^{-1}}$), with the dashed ellipse defined by Simon et
al. (2006a) (based on MSX images), the crosses showing mm emission
cores (Rathborne et al. 2006), and the horizontal line showing a scale
of 3\arcmin and the intensity scale in $\rm MJy\:sr^{-1}$. {\it Middle
left}: Large-scale Median Filter (LMF) estimate of $I_{\rm \nu,0,obs}$
(linear intensity scale in units of ${\rm MJy\:sr^{-1}}$). {\it Bottom
left}: LMF estimate of $\Sigma$ (intensity scale in units of $\rm
g\:cm^{-2}$). {\it Top right}: $I_{\rm \nu,0,obs,LMF}$/$I_{\rm
\nu,0,obs,SMF}$. {\it Middle right}: Small-scale Median Filter (SMF)
estimate of $I_{\rm \nu,0,obs}$. {\it Bottom right}: SMF estimate of
$\Sigma$.
\label{fig:method1}}
\end{center}
\end{figure*}

\subsubsection{Large-Scale Median Filter (LMF)}\label{S:m1}

A relatively simple way of estimating the diffuse IR background at a
given location behind an IRDC is to take a median average of a region
(i.e. filter) centered on the location of interest and that is large
compared to the cloud. This method was applied by Simon et al. (2006a)
to model the Galactic background from {\it MSX} images to then
identify IRDCs as high contrast features. This method will only
capture background fluctuations on scales larger than the
cloud. Indeed if the filter becomes too small, i.e. with a size
comparable to the cloud, then the derived background will become
influenced by the cloud itself and will be underestimated.

\begin{figure}
\begin{center}
\includegraphics[width=\columnwidth]{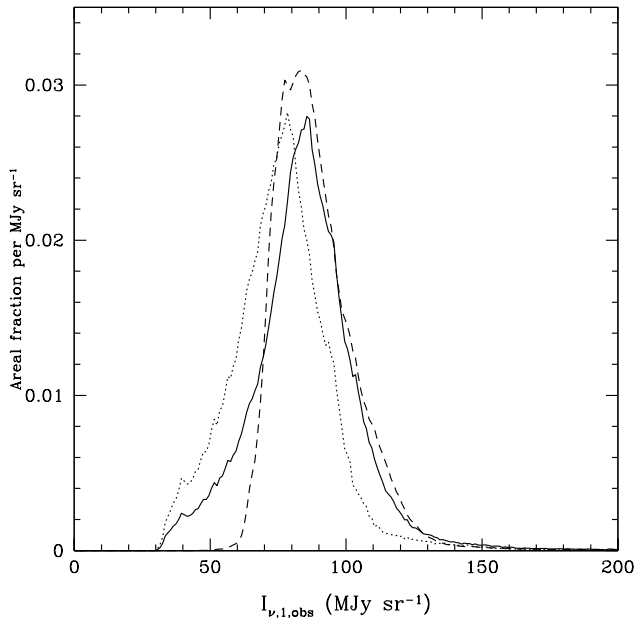}%idistr.eps
\caption{Distribution of pixel intensities for Cloud C. The solid line shows the distribution inside the LMF (13\arcmin\ by 13\arcmin) centered on the IRDC. The dotted line shows the distribution inside the IRDC ellipse and the dashed line shows it outside the ellipse (for a region 24\arcmin\ square).
\label{fig:pixel_inten}}
\end{center}
\end{figure}

For simplicity, we adopt a square-shaped filter for this method. After
some experimentation and given the sizes of the IRDCs in our sample,
we chose a filter size of 13$\arcmin$, i.e. 650 Spitzer-GLIMPSE image
pixels. We evaluate the background on a uniform grid with spacings of
24\arcsec. At each location, we first consider the full distribution
of pixel intensities, and evaluate the mode. Since the high-end tail
of this distribution, due to extended bright sources and stars, was
typically much larger than the low-end tail, due to the IRDC, we
excluded pixels that had intensities twice the value of the mode, and
then found the median value of the remaining distribution. This value
was assigned to $I_{\rm \nu,0,obs}$ for all the pixels in the
$24\arcsec\ \times 24\arcsec$ square sharing the same center as the
13\arcmin\ filter. This Large Scale Median Filter (LMF) method is
illustrated for cloud C in the left-hand panels of
Figure~\ref{fig:method1}. The distribution of pixel intensities is
shown in Figure~\ref{fig:pixel_inten}.

Note that there will be some regions where the estimated value of
$I_{\rm \nu,1}$ is greater than that of $I_{\rm \nu,0}$. This can arise
because of foreground or background stars, or because of small
fluctuations in the true background intensity that are not captured by
the large scale averaging used in the model. Examining the
distribution of intensities in the LMF, we find the FWHM of the
distribution. We set those pixels with $I_{\nu,1}>I_{\nu,0} + 0.5 {\rm
FWHM}$, equal to $I_{\nu,0}$ so that the derived $\Sigma$ at this
location is zero. These regions are typically stars or bright emission
regions. The remaining pixels with $I_{\nu,1}>I_{\nu,0}$ are allowed
to yield a negative value of $\Sigma$, which helps prevent small scale
fluctuations (which can include instrument noise) from biasing the
total mass in a region to positive values.

\subsubsection{Small-Scale Median Filter (SMF)}\label{S:m2}

In order to resolve smaller-scale variations in the IR background, but
without having the median filter estimate be affected by the IRDC, we
develop a second method, which uses a small-scale median filter in
regions outside a defined cloud boundary and an interpolation scheme
inside this boundary. For cloud boundaries, we utilize the ellipses
defined by Simon et al. (2006a) from their study of MSX images of the
clouds. We set the filter size to be one third of the major axis of
this ellipse. The method of estimating the median pixel intensity used
for the LMF (\S\ref{S:m1}) is repeated for the region outside the
cloud ellipse.

\begin{figure}
\begin{center}
\includegraphics[width=\columnwidth]{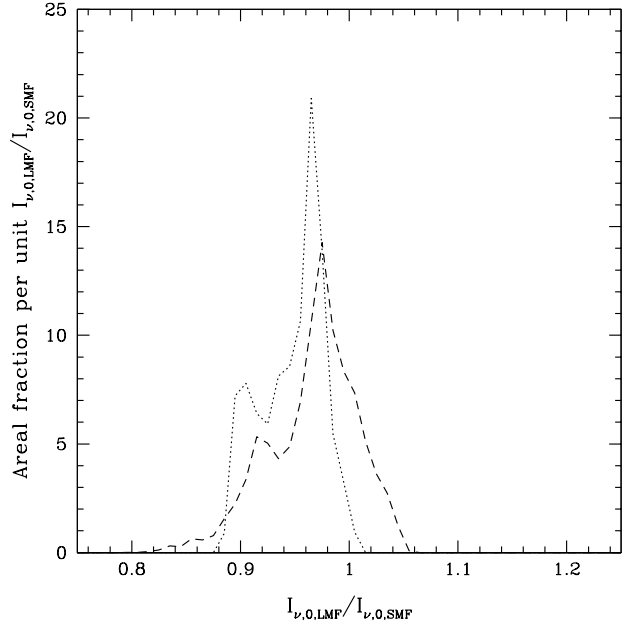}%rdistr.eps
\caption{Distribution of the ratio ($I_{\rm \nu,0,LMF}/I_{\rm
\nu,0,SMF}$) of background intensities derived from of LMF and SMF methods. The dotted and dashed lines show the distribution inside and outside (for a region 24\arcmin\ square) the cloud ellispse, respectively.
\label{fig:ratio}}
\end{center}
\end{figure}

For each image pixel inside the IRDC ellipse we estimate $I_{\rm
\nu,0,obs}$ by interpolating from values of $I_{\rm \nu,0,obs}$
outside of the ellipse that are within an angular distance equal to
the semi-major axis of the ellipse, weighting by the inverse square of
the angular separation, so that the innermost annuli dominate the
average. This Small-Scale Median Filter (SMF) method is illustrated
for cloud C in the right-hand panels of
Figure~\ref{fig:method1}. Here, and also in Figure~\ref{fig:ratio}, we
compare the LMF and SMF estimates of $I_{\nu,0}$, finding typical
variations at the level of $\lesssim 10\%$. A 10\% uncertainty in
$I_{\nu,0}$ corresponds to a mass surface density of $0.013\:{\rm
g\:cm^{-2}}$. We conclude that systematic uncertainties in background
estimation set a minimum threshold $\Sigma$, below which the mid-IR
extinction mapping method becomes unreliable.

Here, we also note that the GLIMPSE images are occasionally prone to
certain artifacts, e.g. changes in diffuse intensity along diagonal
bands and stripes (see cloud B and D images below), which will
introduce additional uncertainties in these regions. The SMF method
improves upon the LMF method in reducing the effect of these artifacts.

\subsubsection{Orthogonal Strips Across Filamentary IRDCs}\label{S:m3}

\begin{figure}
\begin{center}
\includegraphics[width=\columnwidth]{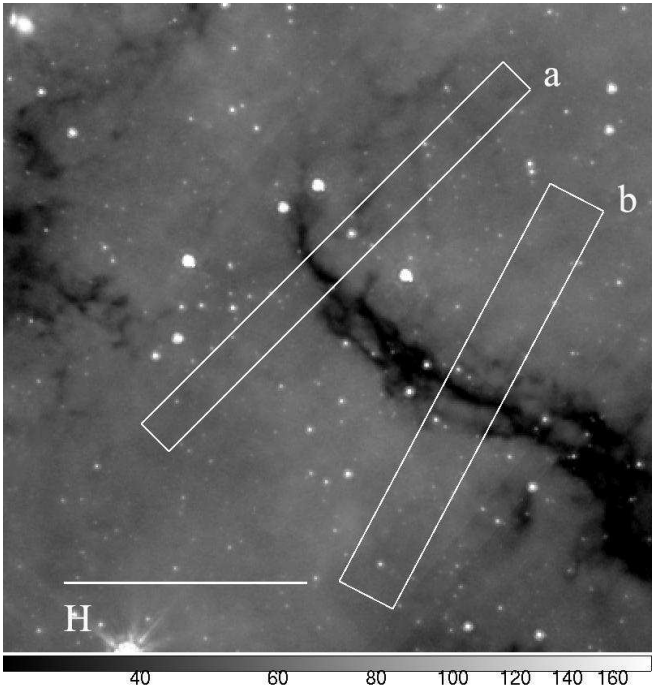}%newstripimage.eps
\caption{GLIMPSE $8\mu$m image of Cloud H with 3\arcmin\ scale bar showing the strips considered in \S\ref{S:m3} (logarithmic intensity scale in units of ${\rm
MJy\:sr^{-1}}$).  
\label{fig:stripimage}}
\end{center}
\end{figure}

\begin{figure*}
\begin{center}
\includegraphics[width=6in]{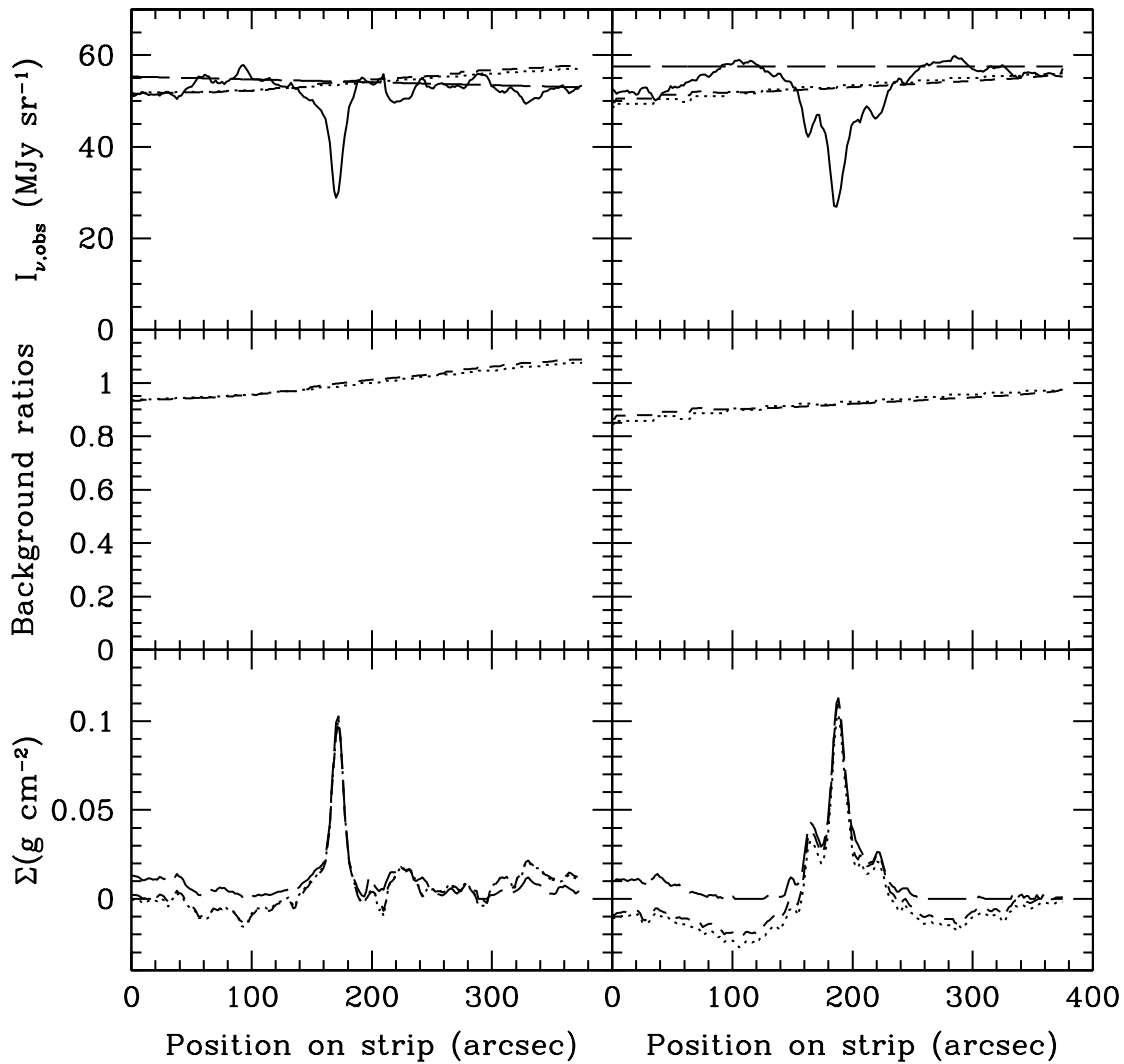}%strip.eps
\caption{Results of LMF and SMF methods with thin strip analyses for
strip a (left column) and strip b (right column). {\it Top} panels: Median
$I_{\rm \nu,1,obs}$ along the strip (solid lines), LMF estimate of
$I_{\rm \nu,0,obs}$ (dotted lines), SMF estimate of $I_{\rm
\nu,0,obs}$ (dashed lines), estimate from strip intensity profile
(long dashed lines). Position coordinate increases from lower-left end
of strips (see Fig.~\ref{fig:stripimage}). {\it Middle} panels: Ratio of LMF
(dotted) and SMF (dashed) estimates of $I_{\rm \nu,0,obs}$ compared to
estimate from the strip intensity profiles. {\it Bottom} panels: Estimates of
$\Sigma$ via LMF (dotted), SMF (dashed), and strip analysis (long
dashed) methods.
\label{fig:strip}}
\end{center}
\end{figure*}

To further test the accuracy of the LMF and SMF methods, we consider
the structure of the IR background on very small scales across a
filamentary IRDC (cloud H). We choose two strips, approximately
orthogonal to regions of the IRDC that are very filamentary
(i.e. thin) and free of stars (see Figure~\ref{fig:stripimage}).  The
median image intensities are evaluated along these strips
(Figure~\ref{fig:strip}). A linear fit for the IR background is fit to
regions of the strip judged to be free of cloud material. This is a
somewhat subjective process, since it is hard to distinguish between
small-scale background variations and additional absorbing
components. Nevertheless, we expect this estimate of the background
intensity just behind the filament to be more accurate than the LMF
and SMF methods, because the interpolation behind the filament is over
a very small angular scale (about a few tens of arcseconds) and the
background appears to be quite smooth in this general region.

The ratio of the LMF and SMF estimates of the background across the
strips compared to that directly estimated from the strips is shown in
Figure~\ref{fig:strip}. We also show the derived column densities. We
see that the variations in the background estimates are relatively
small, at approximately the 10\% level, similar to the variations
between the LMF and SMF methods seen in cloud C, and again
corresponding to mass surface density uncertainties of $\simeq
0.013\:{\rm g\:cm^{-2}}$.

%The mean column density along the strip was calculated for all three
%methods, and the ratio of the backgrounds from \ref{S:m1} and
%\ref{S:m2} to this new background was taken, giving us a reliable
%estimate of the accuracy of our methods(\ref{fig:method3results}).

%Considering only the data inside this rectangle, a series of lines
%parallel to the filament were drawn.  The number of lines and the
%spacing between them was chosen such that each line corresponded to a
%single line of pixels, giving each line a unique set of data points.

%\begin{figure}\label{fig:method3results}
%\begin{center}$
%\begin{array}{ccc}
%\includegraphics[width=3.2in]{medianchangeerrors2f1.ps} & \includegraphics[width=3.2in]{medianchangeerrors2f.ps}
%\end{array}$
%\end{center}
%\caption{Graphs of the median intensities calculated along each of the lines parallel to the strip, described in \ref{S:m3}, for strips "a" and "b", where strip "a" is on the left and "b" is on the right.  The error bars represent the standard deviation of the pixels along the lines mentioned previously. }
%\end{figure}

\section{Results}

\subsection{Mass Surface Density Maps and Distributions}

\begin{figure*}
%newcloudA.eps
%newm2cloudAsigma.eps
%newcloudB.eps
%newm2cloudBsigma.eps
%newcloudD.eps
%newm2cloudDsigma.eps
\begin{center}
%astroph
\includegraphics[width=5in]{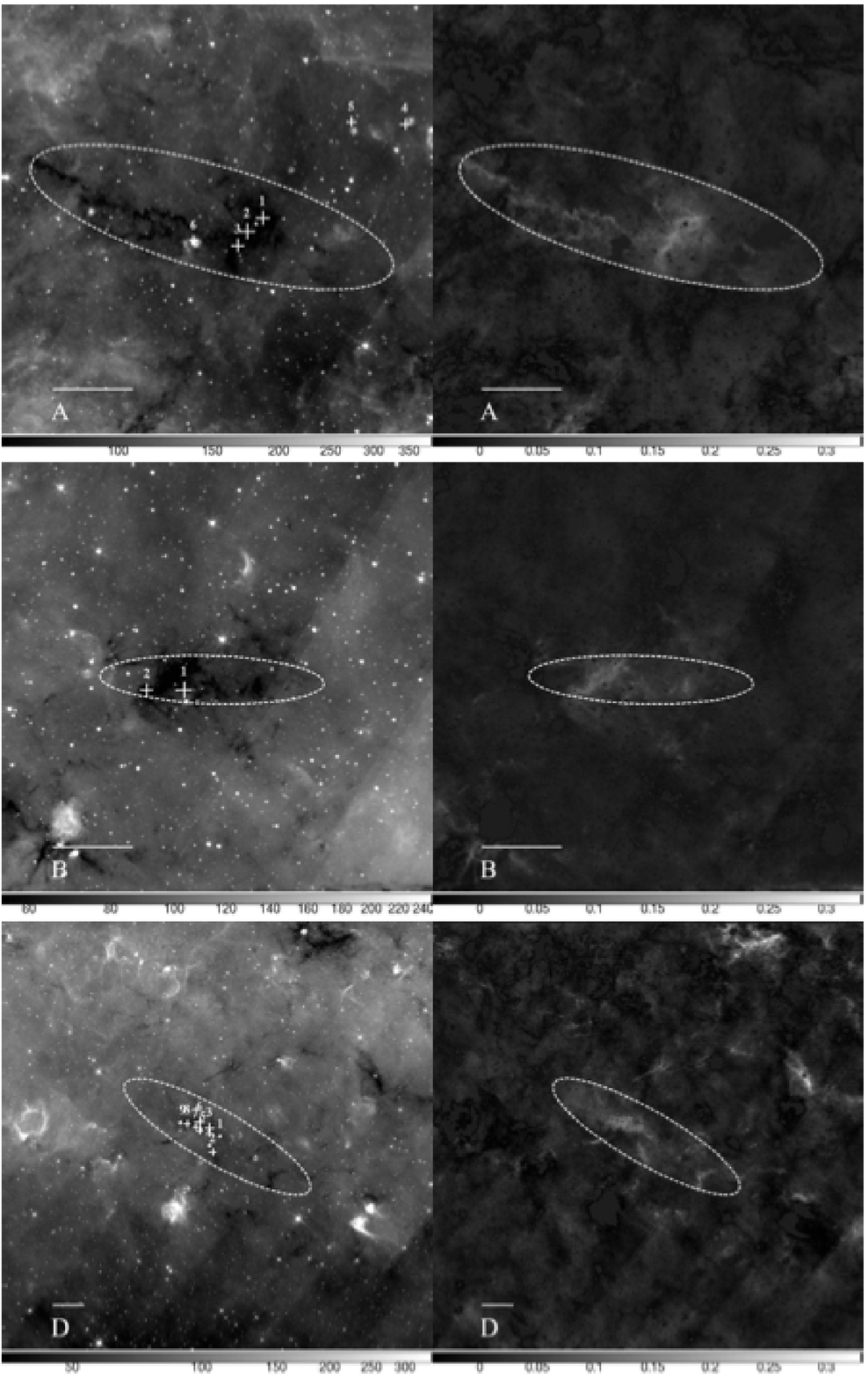}
%$
%\begin{array}{ccc}                         
%\includegraphics[width=2.6in]{f7a_s.eps}
%\includegraphics[width=2.6in]{f7b_s.eps}
% \\
%\includegraphics[width=2.6in]{f7c_s.eps}
%\includegraphics[width=2.6in]{f7d_s.eps}
%\\
%\includegraphics[width=2.6in]{f7e_s.eps}
%\includegraphics[width=2.6in]{f7f_s.eps}
%\end{array}$
\caption{\footnotesize The {\it left} column shows Spitzer IRAC 8$\mu m$ images of
IRDCs, with the dashed ellipse defined by Simon et al. (2006a) (based
on MSX images), the crosses showing mm emission cores (Rathborne et
al. 2006), with the numbers referring to their designation in
Table~\ref{tb:cores}, and the horizontal line showing a scale of
3\arcmin. The {\it right} column shows extinction maps made by the
small-scale median filtering method (SMF). From {\it top} to {\it
bottom}, the clouds are: Cloud A (G018.82$-$00.28), Cloud B
(G019.27$+$00.07), Cloud D (G028.53$-$00.25).
\label{fig:results1}                          }
\end{center}                               
\end{figure*}                               

\begin{figure*}
%newcloudE.eps
%newm2cloudEsigma.eps
%newcloudF.eps
%newm2cloudFsigma.eps
%newcloudG.eps
%newm2cloudGsigma.eps
\begin{center}
%astroph
\includegraphics[width=5in]{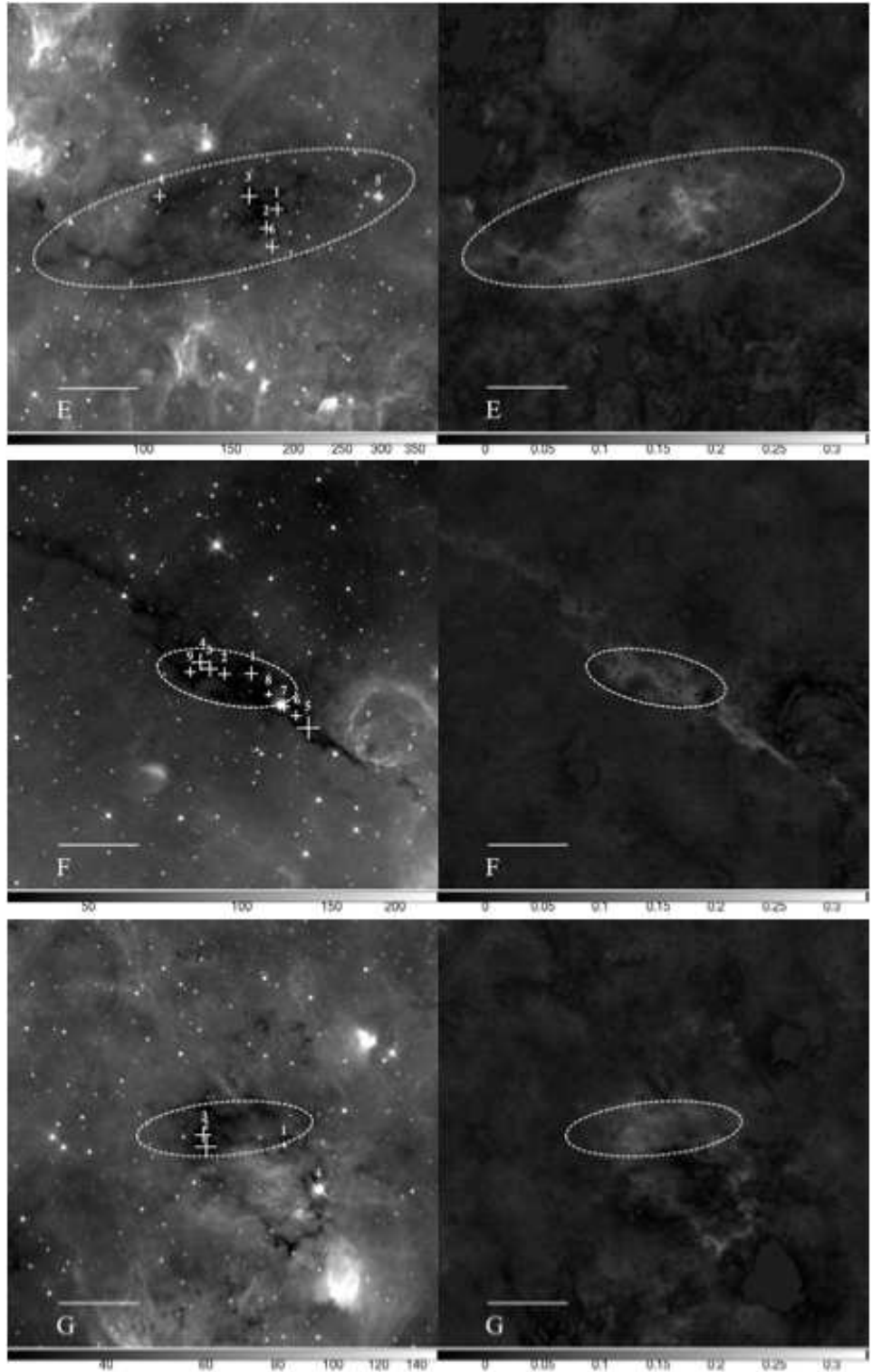}
%$
%\begin{array}{ccc}                         
%\includegraphics[width=2.6in]{f8a_s.eps}
%\includegraphics[width=2.6in]{f8b_s.eps}
% \\
%\includegraphics[width=2.6in]{f8c_s.eps}
%\includegraphics[width=2.6in]{f8d_s.eps}
%\\
%\includegraphics[width=2.6in]{f8e_s.eps}
%\includegraphics[width=2.6in]{f8f_s.eps}
%\end{array}$
\caption{Same as Figure~\ref{fig:results1}, but now, from {\it top} to
{\it bottom}, the clouds are: Cloud E (G028.67$+$00.13), Cloud F
(G034.43$+$00.24), Cloud G (G034.77$-$00.55).
%The {\it left} column shows Spitzer IRAC 8$\mu m$ images of
%IRDCs, with the dashed ellipse defined by Simon et al. (2006a) (based
%on MSX images), the crosses showing mm emission cores (Rathborne et
%al. 2006), and the horizontal line showing a scale of 3\arcmin. The
%{\it right} column shows extinction maps made by the small-scale
%median filtering method (SMF). 
\label{fig:results2} }
\end{center}                               
\end{figure*}        

\begin{figure*}
%newcloudH.eps
%newm2cloudHsigma.eps
%newcloudI.eps
%newm2cloudIsigma.eps
%newcloudJ.eps
%newm2cloudJsigma.eps
\begin{center}
\includegraphics[width=5in]{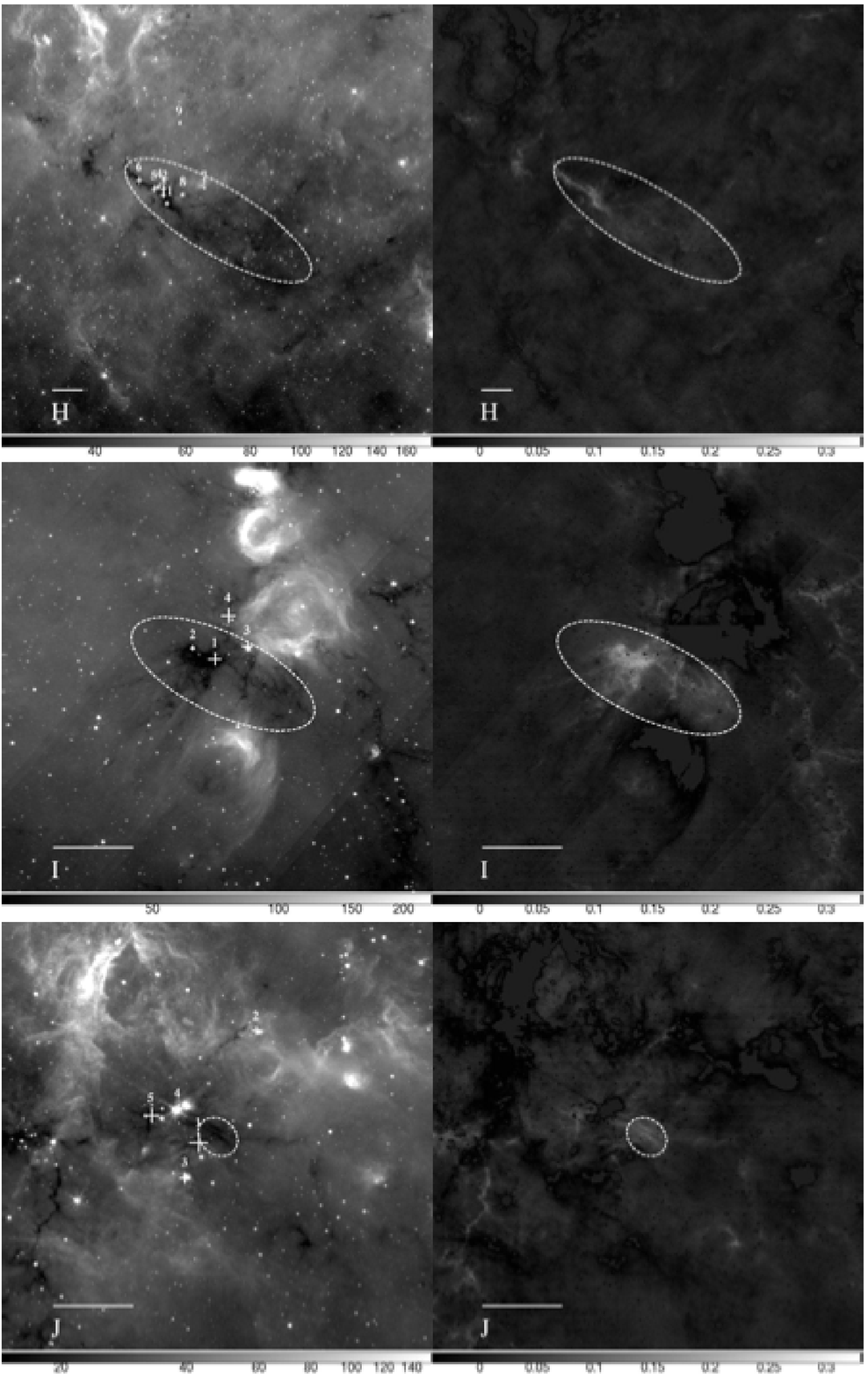}
%$
%\begin{array}{ccc}                         
%\includegraphics[width=2.6in]{f9a_s.eps}
%\includegraphics[width=2.6in]{f9b_s.eps}
% \\
%\includegraphics[width=2.6in]{f9c_s.eps}
%\includegraphics[width=2.6in]{f9d_s.eps}
%\\
%\includegraphics[width=2.6in]{f9e_s.eps}
%\includegraphics[width=2.6in]{f9f_s.eps}
%\end{array}$
\caption{Same as Figure~\ref{fig:results1}, but now, from {\it top} to {\it bottom}, the
clouds are: Cloud H (G035.39$-$00.33), Cloud I (G038.95$-$00.47),
Cloud J (G053.11$+$00.05).
%The {\it left} column shows Spitzer IRAC 8$\mu m$ images of
%IRDCs, with the dashed ellipse defined by Simon et al. (2006a) (based
%on MSX images), the crosses showing mm emission cores (Rathborne et
%al. 2006), and the horizontal line showing a scale of 3\arcmin. The
%{\it right} column shows extinction maps made by the small-scale
%median filtering method (SMF). From {\it top} to {\it bottom}, the
%clouds are: Cloud H (G035.39$-$00.33), Cloud I (G038.95$-$00.47),
%Cloud J (G053.11$+$00.05).
\label{fig:results3}}
\end{center}                               
\end{figure*}

The mass surface density maps derived using the SMF method for the
remaining 9 clouds are presented in
Figures~\ref{fig:results1}---\ref{fig:results3} with a uniform scale
range in $\Sigma$. These IRDCs exhibit a variety of morphologies,
ranging from very filamentary (clouds F and H) to those with more
apparently spherical distributions (clouds C and E). The derived mass
surface densities range up to $\simeq 0.35\:{\rm g\:cm^{-2}}$, which
is likely to be dependent on the angular resolution of the images. As
is apparent from the 8~$\rm \mu$m images, some of the clouds have a
few apparently embedded sources, which cause localized regions in the
clouds to be IR-bright, and thus not amenable to our extinction
mapping technique.

\begin{figure}
\begin{center}
\includegraphics[width=\columnwidth]{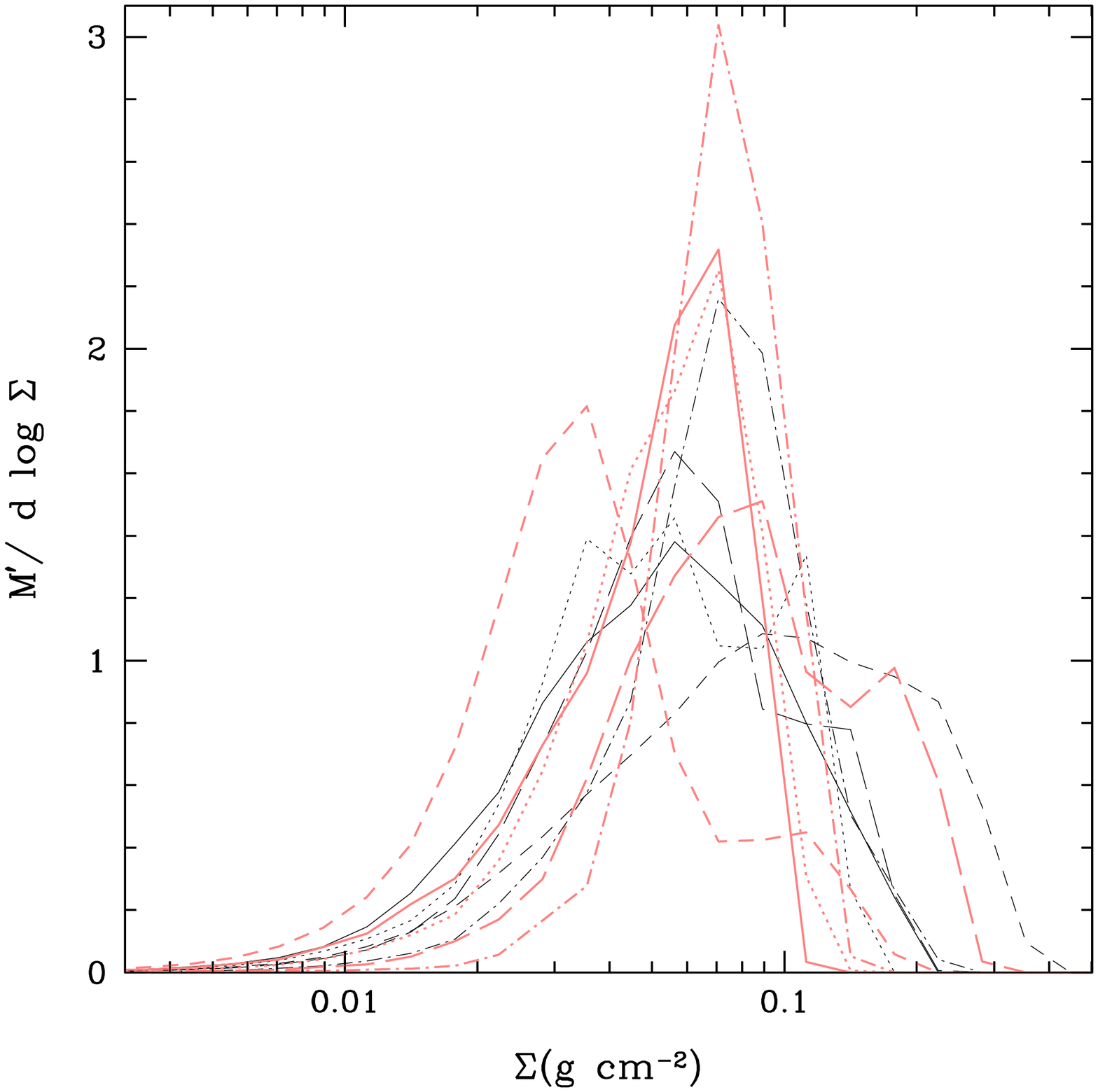}%sigdistrnew.eps
\caption{Mass-weighted probability distribution function ($M^\prime$
is the mass fraction) of $\Sigma$ for clouds A (thin solid black
line), B (thin dotted black), C (thin dashed black), D (thin
long-dashed black), E (thin dot-dashed black), F (thick solid gray
[pink in electronic version]), G (thick dotted gray), H (thick dashed
gray), I (thick long-dashed gray), J (thick dot-dashed gray).
\label{fig:sigdistrnew}}
\end{center}
\end{figure}

\begin{figure}
\begin{center}
\includegraphics[width=\columnwidth]{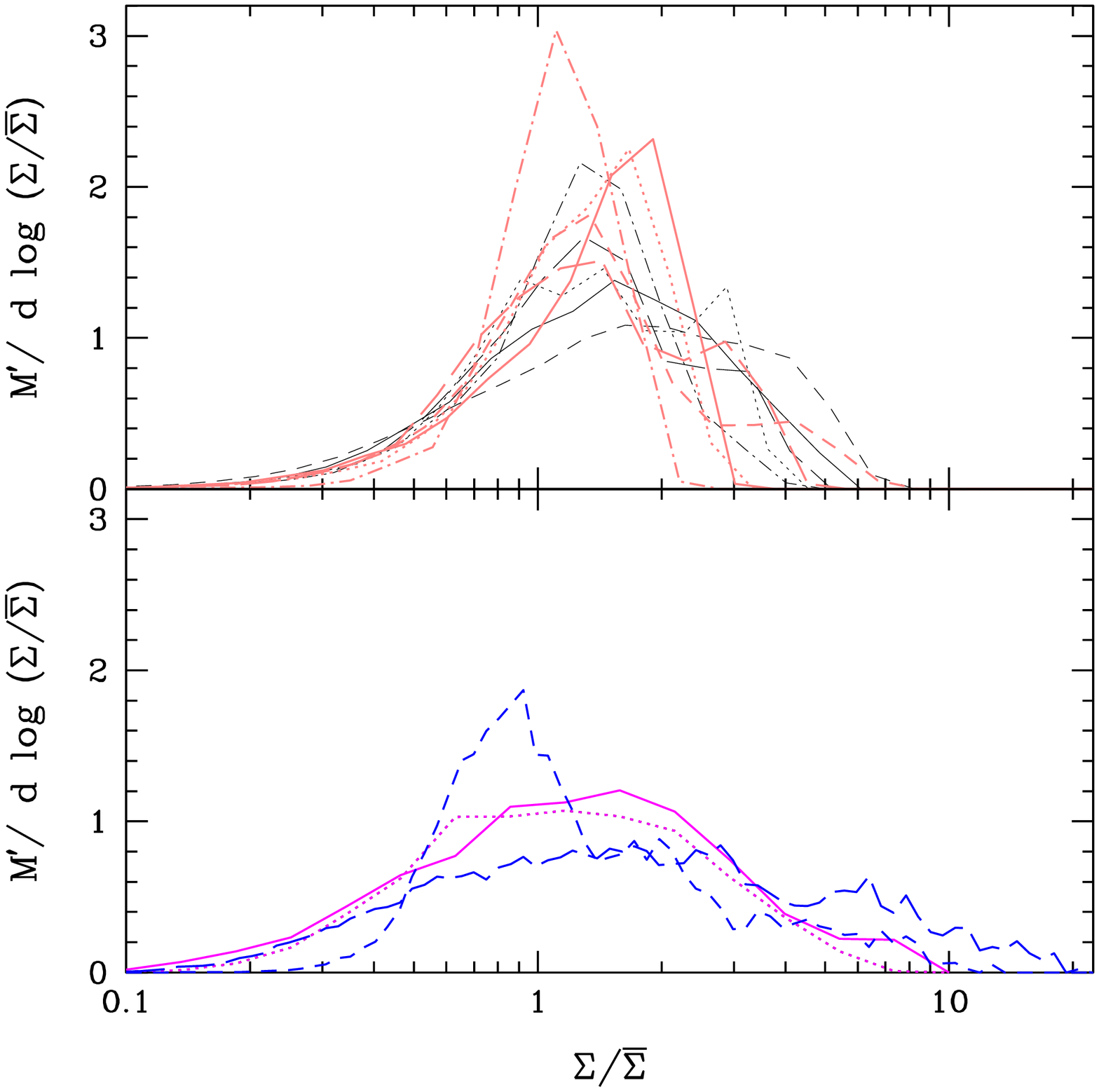}%sigdistrnormnew2.eps
\caption{{\it Top}: Mass-weighted probability distribution function of
$\Sigma/\bar{\Sigma}$ for clouds A (thin solid black line), B (thin
dotted black), C (thin dashed black), D (thin long-dashed black), E
(thin dot-dashed black), F (thick solid gray [pink in electronic
version]), G (thick dotted gray), H (thick dashed gray), I (thick
long-dashed gray), J (thick dot-dashed gray).  Note, cloud J has the
most sharply-peaked, narrowest distribution, while cloud C has the
broadest.  {\it Bottom}: Comparison to simulations of astrophysical
turbulence. Dotted line shows the result of Offner et al. (2008) for
driven (Mach 4.7) turbulence with no self-gravity. Solid line shows
the same simulation after self-gravity has been allowed to operate for
1 mean free fall time. The long-dashed line shows the result of
Nakamura \& Li (2007) for protostellar driven turbulence after 1.5
global gravitational collapse times as an average of two orthogonal
viewing angles that are both perpendicular to the mean large-scale
B-field in the simulation. The dashed line shows the view along the
mean B-field.
\label{fig:sigdistrnormnew}}
\end{center}
\end{figure}

In figure \ref{fig:sigdistrnew} we show the mass-weighted probability
distribution functions (PDFs) with $\Sigma$ (evaluated with the SMF
method) of the 10 IRDCs using the regions inside their Simon et
al. (2006a) ellipses. Here we define $M^\prime$ as the mass fraction,
normalized over the distribution of pixels with $\Sigma>0$. In figure
\ref{fig:sigdistrnormnew} we show the same PDF but now as a function
of $\Sigma/\bar{\Sigma}$, where $\bar{\Sigma}$ is the mean $\Sigma$
(area-weighted over pixels which have $\Sigma>0$). The values of
$\bar{\Sigma}$ are listed in Table~\ref{tb:clouds}.

The distributions of $\Sigma/\bar{\Sigma}$ show the most similarity
for $\Sigma/\bar{\Sigma}<0.5$. 
%However, recall the regions of lower $\Sigma$ are prone to the largest uncertainties. 
The high-side of the distributions do show a wide variety of profiles:
there is a large range of mass fractions at mass surface densities
that are, for example, 3 times greater than the mean. Cloud J has the
narrowest PDF, while cloud C has the broadest and one of the most
skewed distributions.  We caution that our derived IRDC PDFs are
likely to be sensitive to the geometry of the ellipse that was chosen
to represent the IRDC. For example this can affect the normalization
of $\bar{\Sigma}$ (although note we restrict its calculation over only
the area of the cloud for which we measure $\Sigma>0$).
%that they have the largest mass fractions at
%$\Sigma/\bar{\Sigma}\simeq 3$. Several of the IRDCs exhibit a
%double-peaked PDF.
%I have the broadest. The latter
%clouds appear, qualitatively at least, to be surrounded by relatively
%strong 8~$\rm \mu m$ emission features, which might indicate enhanced
%injection of turbulent energy into the gas. 

In the lower panel of Figure~\ref{fig:sigdistrnormnew} we make a
qualitative comparison of these results with the shapes of $\Sigma/\bar{\Sigma}$
PDFs resulting from different astrophysical sources of turbulence as
modeled in two sets of published numerical simulations. Offner, Klein,
\& McKee (2008) presented simulations of isothermal driven turbulence
to maintain a mean 1D Mach number of about 5 in which self-gravity was
then turned on at a particular time. One expects clouds with higher
Mach number turbulence to have broader PDFs of $\Sigma/\bar{\Sigma}$,
as stronger shocks lead to greater contrasts between regions of
compression and rarefaction. Clouds that are more strongly
self-gravitating also have broader PDFs. Self-gravity also tends to
skew the PDFs towards the high-value side.
Figure~\ref{fig:sigdistrnormnew} shows two examples from these
simulations: one is before self-gravity has been initiated, so is
representative of the $\Sigma/\bar{\Sigma}$ PDF from hydrodynamic Mach
4.7 (1D) driven turbulence with a Burgers $P(k)\propto k^{-2}$ power
spectrum; the other, which has a similar Mach number of 4.9, is after
self-gravity has been allowed to develop structures for about 1 mean
simulation free-fall time while the turbulence is still being driven
and is somewhat broader as a result (14.2\% of the gas mass has
collapsed into sink particles by this time; the PDF is normalized to
the total remaining gas mass).

Nakamura \& Li (2007) presented simulations of magnetized protostellar
outflow driven turbulence. The driving scale is relatively small
compared to the simulation box, and the sources are centrally
concentrated in the box. Their formation is sporadic as the global
clump collapses: we show results after 1.5 global gravitational
collapse times, when about 80 protostars have formed. Furthermore
there is a large scale direction to the initial magnetic field, which
has a mean dimensionless flux-to-mass ratio of 0.52 (in units of the
critical value, $2\pi G^{1/2}$ [Nakano \& Nakamura 1978]), decreasing
to 0.19 in the clump center. The mean value corresponds to about
$75\:{\rm \mu G}$ for the fiducial simulation parameters. This inhibits
perpendicular motions, as can be seen from the narrowness of the
$\Sigma/\bar{\Sigma}$ PDF derived looking along the field direction
compared to perpendicular to it (Figure~\ref{fig:sigdistrnormnew};
these distributions correspond to the models shown in Figure 7 of
Nakamura \& Li [2007]).

A comparison of the observed and simulated $\Sigma/\bar{\Sigma}$ PDFs
shows that the ensemble of the observed distributions can be
qualitatively accounted for by the range of simulations shown,
although the observed IRDCs generally have narrower
distributions. However, we are not able to exclude the possibility
that other numerical models involving different physics, e.g. a larger
degree of magnetic support, and/or other parameters could not also
explain the observed clouds. Since most of the IRDCs have relatively
narrow PDFs compared to the Mach 5 turbulence models, this may
indicate that these clouds have smaller turbulent velocity dispersions
and/or that dynamically important magnetic fields are present. The
protostellar outflow driven turbulence models generally have broader
$\Sigma/\bar{\Sigma}$ PDFs than the observed IRDCs, and this may
indicate these models are more applicable to later evolutionary stages
of star cluster formation. However, they do illustrate the effects of
dynamically important large-scale magnetic fields in creating narrower
$\Sigma/\bar{\Sigma}$ PDFs for viewing directions along the field
lines. 

Observations of the velocity structure of these IRDCs as traced by
molecular line emission can help to further test the nature of
turbulence present in the early stages of star cluster formation.  The
extinction mapping technique we have presented also opens up the
possibility of studying the $\Sigma/\bar{\Sigma}$ PDF as a function of
IRDC $\bar{\Sigma}$, $M$ and embedded stellar content.  We defer a
more quantitative study of these dependencies and a quantitative
comparison of the $\Sigma/\bar{\Sigma}$ PDFs to those formed in
numerical simulations to a future study.

\subsection{Cloud and Core Masses}

Given the (kinematic) distance to each cloud (Simon et al. 2006a), we
convert our $\Sigma$ estimates into mass estimates. For a typical
distance of 3~kpc, each 2.0\arcsec\ resolution element at $\rm 8\:\mu
m$ (i.e. the mean FWHM of the point response function) corresponds to
a linear scale of 0.029~pc, and if this has a typical $\Sigma$ of
0.1~$\rm g\:cm^{-2}$ , this corresponds to a mass of
0.41~$M_\odot$. We see that the mid-IR extinction mapping technique has the
potential to probe relatively low mass scales.

In Table~\ref{tb:clouds} we list the total cloud masses derived by the
LMF and SMF extinction mapping methods. Note, these are the masses
inside the elliptical regions defined by Simon et al. (2006a) from MSX
images, and thus are not necessarily particularly close
representations of the morphologies revealed by {\it Spitzer} IRAC. We find a
mean fractional difference of 35\% between these two mass
estimates. We regard the SMF mass estimate as being more accurate.
Uncertainties in both methods will grow for larger clouds with lower
mean mass surface densities (i.e. smaller contrasts against the
background). Note also this extinction mapping technique is not
sensitive to a uniform screen of matter that covers both the region of
the IRDC and the region where the background is estimated. A
comparison of extinction-derived cloud masses and mass surface
densities with the properties inferred by molecular line emission,
such as $\rm ^{13}CO$, will be presented in a separate study (A. K. Hernandez \&
J. C. Tan, in prep.).

The extinction mass estimates are expected to become more accurate at
high values of $\Sigma$. Each of the IRDCs in our sample contains
dense cores studied by their mm dust continuum emission by Rathborne
et al. (2006). We first identify those cores which are amenable to
extinction mapping (i.e. do not contain bright $8\:{\rm \mu m}$
emission). Cores identified by Rathborne et al. (2006) as being mid-IR
bright (labelled ``(e)'' in Table \ref{tb:cores}) were excluded. We
also excluded cores overlapping with fainter IR sources, if the area
affected by the sources was greater than about 10\% of the
core. Finally, we excluded cores with very low surface densities,
$\Sigma< 0.02\: {\rm g\:cm^{-2}}$, since the extinction mapping method
becomes very unreliable at these levels due to uncertainties in the
background estimation.

For our selected cores, we summed the mass inside a circular area
centered on each core position with radius equal to half the angular
FWHM diameter of Gaussian fits that Rathborne et al. (2006) fitted to
their mm continuum images. Note the quoted angular FWHM diameters are
typically about a factor of 1.09 times larger than the deconvolved
FWHM diameters, so our areal integration actually extends out to about
0.545 true FWHM diameters from the core center. The total core masses
quoted by Rathborne et al., which are evaulated by integrating over
the whole 2D Gaussian distribution, will thus be a factor of 1.78
larger than that contained in this angular area, so we reduce their
masses by a factor of 0.562. Their mass estimates involved assuming
$T=15$~K, a dust opacity per mass of dust of $\kappa_{\rm 1.2mm}
=1.0\:{\rm cm^2\:g^{-1}}$ (Ossenkopf \& Henning 1994), a gas-to-dust
mass ratio of 100, and graybody emission with emissivity index
$\beta=2$. In the context of the Ossenkopf \& Henning (1994) dust
model with thin ice mantles, coagulated for $10^5$~yr at $n_{\rm
H}=10^6\:{\rm cm^{-3}}$, which is the closest to our adopted 8~$\rm
\mu m$ opacity, the interpolated 1.2~mm opacity is $1.056\:{\rm
cm^2\:g^{-1}}$, i.e. 5.6\% higher than assumed by Rathborne et
al. (2006), so we further reduce their masses by a factor 0.947. Our
calculation of $\Sigma$ from dust extinction assumed a gas-to-dust
ratio of 156, which was that adopted for the Ossenkopf \& Henning
(1994) coagulation model. If this same ratio is applied to the mm
emission masses, then the derived masses would be raised by 1.56 from
the Rathborne et al. estimates. These three effects imply the
Rathborne et al. quoted core masses should be reduced by a factor of
0.830 for a fair comparison with our extinction masses. Note, we use
the same distance adopted by Rathborne et al. to the cores. In
Table~\ref{tb:cores} and Figure~\ref{fig:cores} we compare the mm dust
emission and mid-IR dust extinction mass estimates.

We find generally very good agreement between these different methods.
%The standard deviation for method1/emission is 0.235112, and the stddev
%for method1/method2 is 0.0594611, all in log10.
The dispersion in the ratios of the LMF and SMF estimates is about
15\%. Comparing to the mm emission masses, the logarithmic average of
$M_{\rm mm}/M_{\rm SMF}$ is 1.08 with a dispersion of a factor of
1.8. This small systematic offset may be somewhat fortuitous given the
systematic errors inherent to both methods. The possibility of
correlations of $M_{\rm mm}/M_{\rm SMF}$ with core properties is
discussed below.

\subsection{Correlations of Core $M_{\rm mm}/M_{\rm SMF}$ with Density as Evidence for Grain Growth}

\begin{figure*}
\begin{center}
\includegraphics[width=6in]{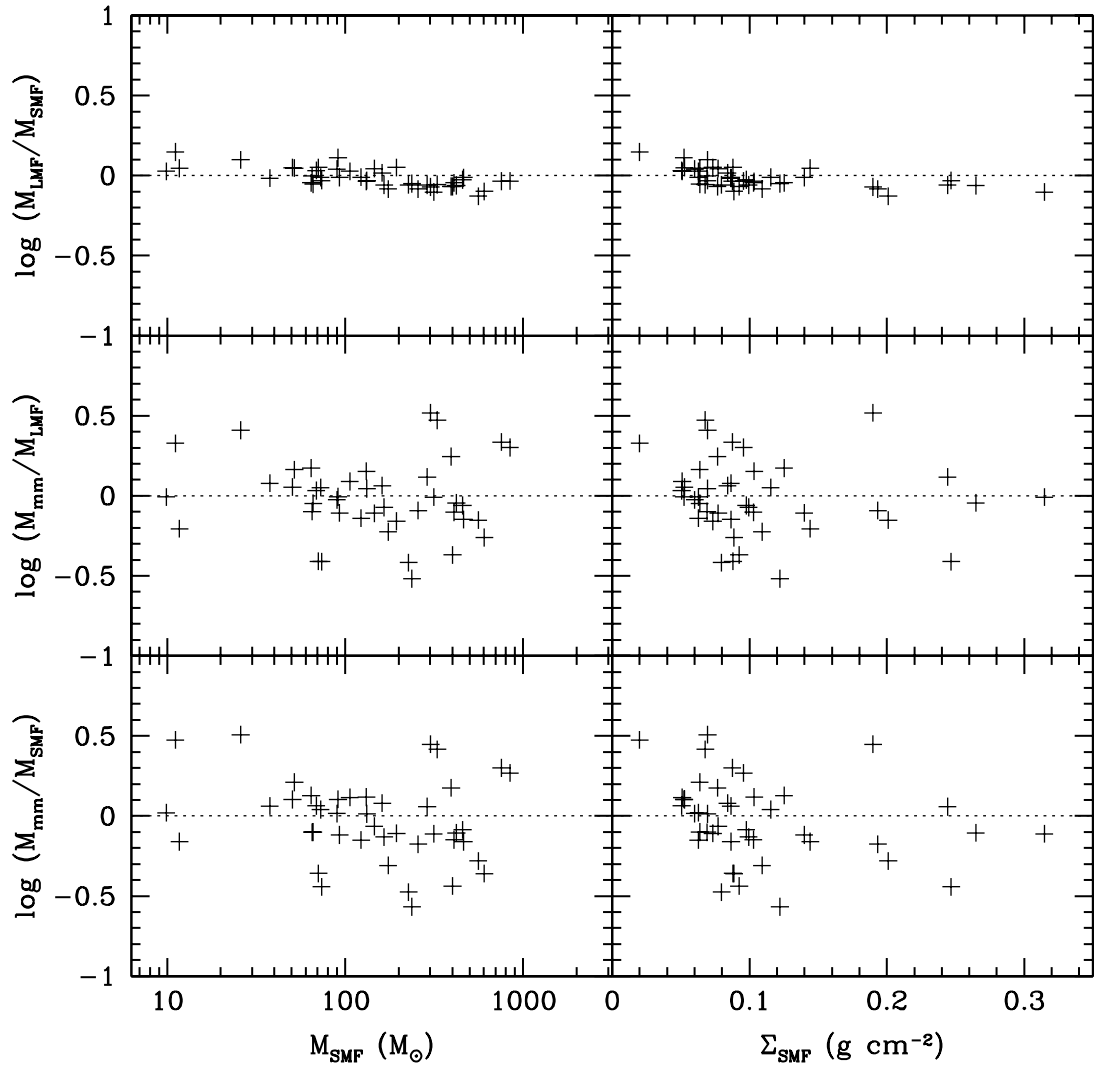}%cores.eps
\caption{Comparison of core mass estimates. {\it Top Left}: Ratio of cores
masses determined via extinction mapping with Large and Small Median
Filter methods ($M_{\rm LMF}/M_{\rm SMF}$) versus core mass. {\it Top
right}: Same data plotted as a function of core $\Sigma$ (estimated via
the SMF method). {\it Middle} panels: Ratio of mm emission mass (Rathborne et
al. 2006) to $M_{\rm LMF}$. {\it Bottom} panels: Ratio of mm emission mass
to $M_{\rm SMF}$.
\label{fig:cores}} 
\end{center}
\end{figure*}

\begin{figure*}
\begin{center}
\includegraphics[width=2.4in]{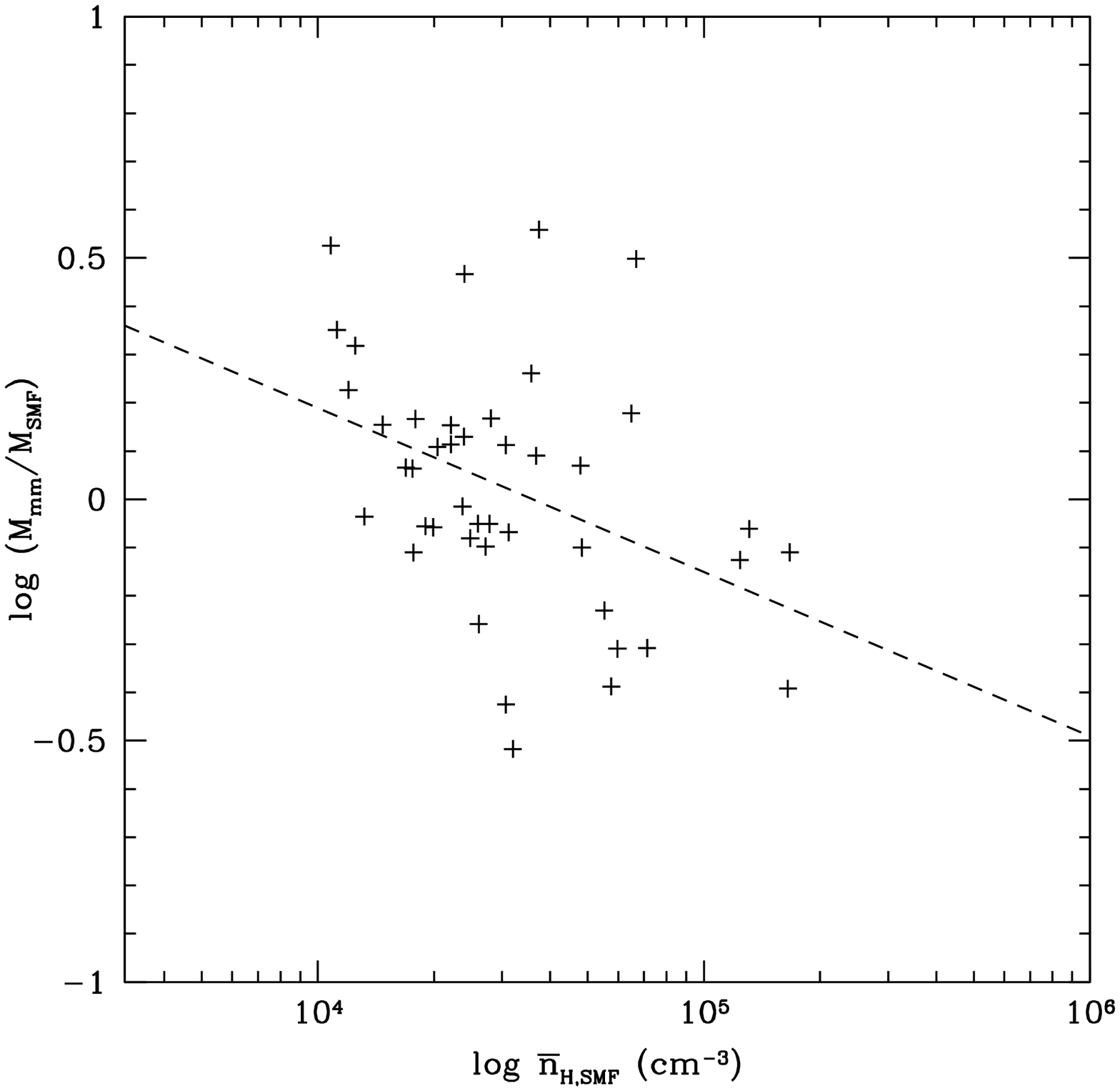}%coresden.eps
\includegraphics[width=2.4in]{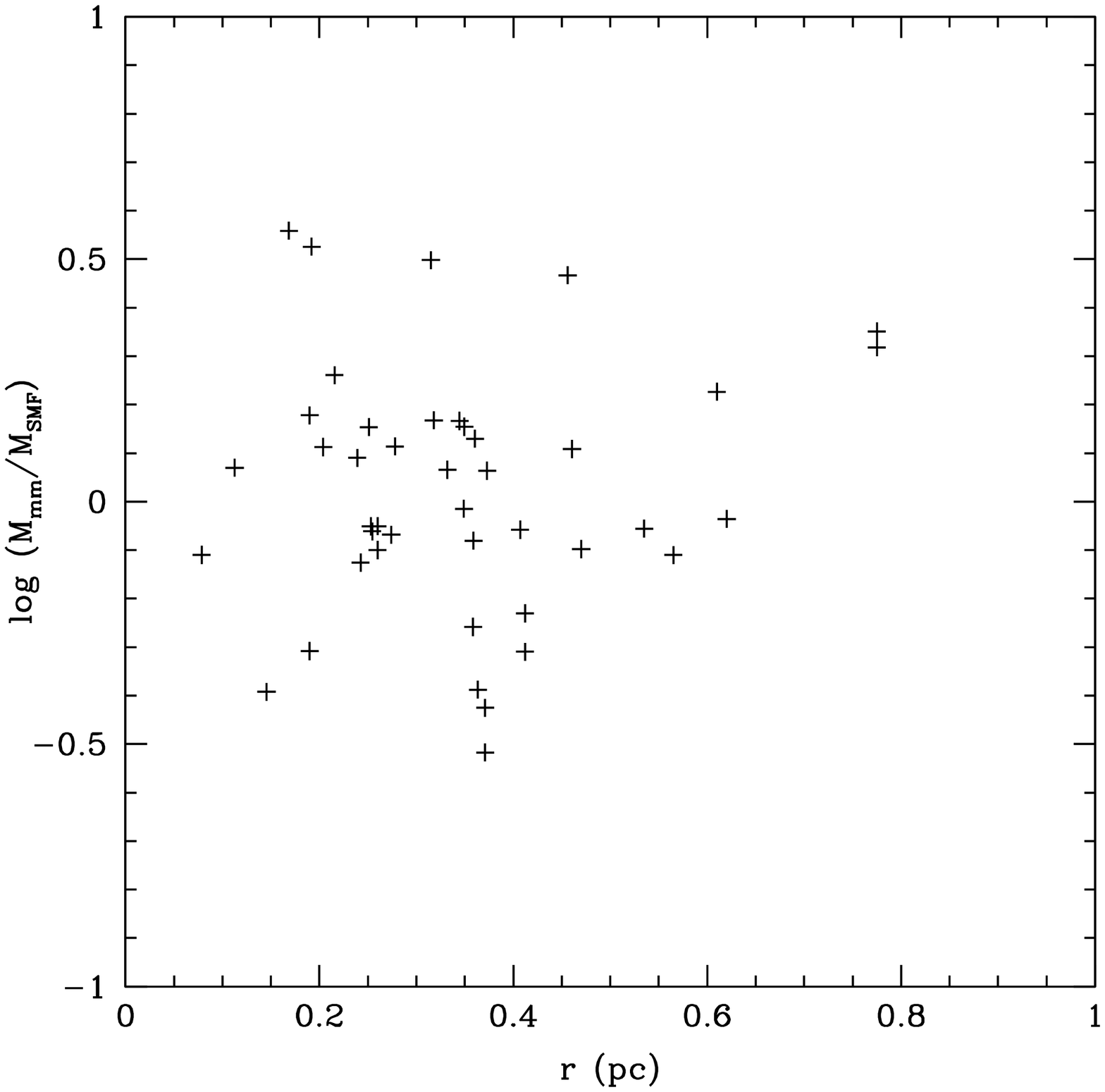}%coresdenr.eps
\includegraphics[width=2.4in]{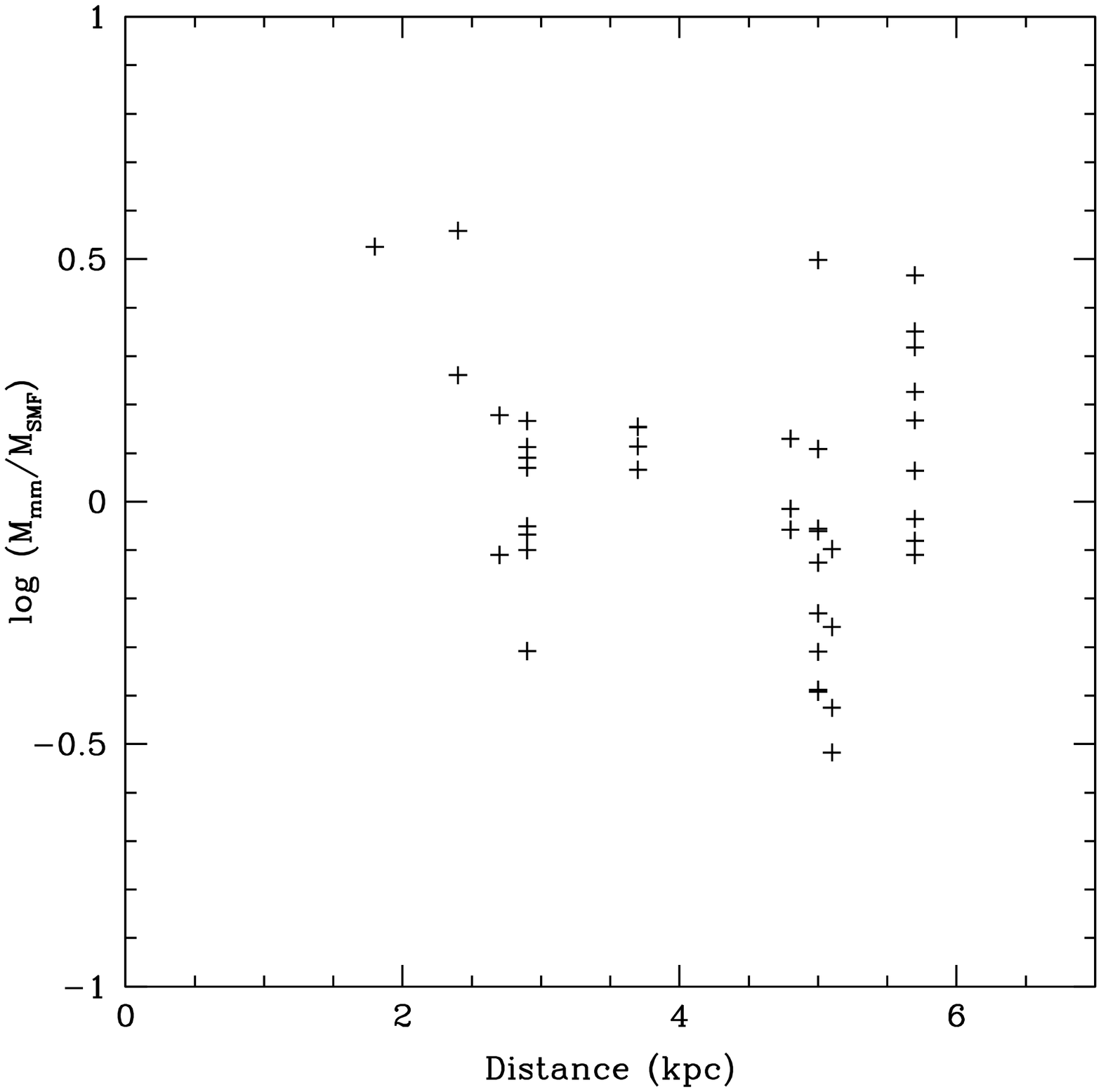}%coresdend.eps
\caption{ {\it Left} panel: Ratio of core masses estimated from mm dust
emission and 8~$\rm \mu$m extinction with the SMF method ($M_{\rm
mm}/M_{\rm SMF}$) versus mean volume density, $\bar{n}_{\rm
H,SMF}$. The dashed line shows the best fit power law relation,
$(M_{\rm mm}/M_{\rm SMF}) \propto \bar{n}_{\rm H,SMF}^{-0.33}$. The
significance of this anticorrelation is discussed in the text. {\it Middle}
panel: Ratio of ($M_{\rm mm}/M_{\rm SMF}$) versus core radius. No
significant trend is observed (see text). {\it Right}
panel: Ratio of ($M_{\rm mm}/M_{\rm SMF}$) versus core distance. No
significant trend is observed (see text).
\label{fig:coresden}} 
\end{center}
\end{figure*}

We also use our SMF extinction mass to calculate a volume-averaged H
number density in each core, $\bar{n}_{\rm H,SMF}$, assuming spherical
geometry. We find values in the range $\bar{n}_{\rm H,SMF}\simeq
10^{4} - 10^5\:{\rm cm^{-3}}$. We examine the ratio of $M_{\rm
mm}/M_{\rm SMF}$ as a function of core density in
Figure~\ref{fig:coresden}. The best fit power law relation is $(M_{\rm
mm}/M_{\rm SMF}) \propto \bar{n}_{\rm H,SMF}^{-0.33}$.  Considering
the Spearman rank-order correlation, the probability of a chance
correlation is $0.0036$. Since $M_{\rm SMF}$ and $\bar{n}_{\rm H,SMF}$
are correlated via the observable $\Sigma_{\rm SMF}$, we also examine
the correlation of $(M_{\rm mm}/M_{\rm SMF})$ with $M_{\rm SMF}$,
$\Sigma_{\rm SMF}$ (Figure~\ref{fig:cores}), core radius, $r$, and
core distance (Figure~\ref{fig:coresden}). We find probabilities of
chance (anti)correlation of 0.28, 0.018, 0.99, and 0.43 respectively,
i.e. there is no significant correlation of $(M_{\rm mm}/M_{\rm SMF})$
with $M_{\rm SMF}$, $r$, and distance, and there is a marginally
significant anticorrelation with $\Sigma_{\rm SMF}$. The most
significant trend is the anticorrelation of $(M_{\rm mm}/M_{\rm SMF})$
with density.

One possible explanation for this observed trend is a systematic
change in dust opacities at $\rm 8 \mu$m and 1.2~mm caused by grain
growth, since $M_{\rm mm}/M_{\rm SMF} \propto \kappa_{\rm 8\mu
m}/\kappa_{\rm 1.2mm}$. For example, the thin ice mantle model of
Ossenkopf \& Henning (1994) predicts $\kappa_{\rm IRAC 8\mu
m}/\kappa_{\rm 1.2mm}$ decreases by a factor of 0.68 after coagulation
for approximately $10^6\:{\rm yr}$ at a density of $n_{\rm
H}=10^5\:{\rm cm^{-3}}$ (or $10^7\:{\rm yr}$ at a density of $n_{\rm
H}=10^4\:{\rm cm^{-3}}$). The ratio decreases only slightly in the
limit of further coagulation. Comparison of the uncoagulated bare
grain and thin ice mantle models of Ossenkopf \& Henning (1994) shows
that $\kappa_{\rm 8\mu m}/\kappa_{\rm 1.2mm}$ decreases by about a
factor of 0.75 by the formation of ice mantles. Thus a total factor of
about 0.5 decrease in $M_{\rm mm}/M_{\rm SMF}$, i.e. about that observed
in our cores as the mean density changes from $\sim 10^4\:{\rm
cm^{-3}}$ to $\sim 10^5\:{\rm cm^{-3}}$, could be explained by a
combination of ice mantle formation and grain growth. Note the typical
density in the cores could be higher than the mean volume averaged
value if they are clumpy.

Systematic temperature changes may also lead to changes in $M_{\rm
mm}/M_{\rm SMF}$, via the estimate of $M_{\rm mm}$. One does expect
colder temperatures at higher densities, which would lead to $M_{\rm
mm}$ being relatively underestimated at high densities under the
constant temperature assumption of Rathborne et al. (2006). For
example, adopting a temperature of 10~K rather than 15~K would
raise $M_{\rm mm}$ by a factor of 1.9, which is enough to account for
the observed trend of $M_{\rm mm}/M_{\rm SMF}$ with density.

We conclude that this IRDC core population, as the mean density
increases from $\sim 10^4\:{\rm cm^{-3}}$ to $\sim 10^5\:{\rm
cm^{-3}}$, shows tentative evidence for either opacity changes that
are consistent with models of ice mantle formation and grain
coagulation, or a temperature decrease of $\sim 5$~K. Systematic
follow up of these cores with $\rm NH_3$ observations to measure
temperature, as has been done in some IRDCs (e.g. Sridharan et
al. 2005; Pillai et al. 2006, Wang et al. 2008), can help to
distinguish these possibilities.

Evidence for grain growth has previously been reported in the low-mass
core B68 by Bergin et al. (2006), based on low gas-to-dust temperature
coupling rates. Bianchi et al. (2003) considered ratios of 850~$\rm
\mu m$ and 1.2~mm opacities to visual opacities in B68, finding no
conclusive evidence for grain growth. Keto \& Caselli (2008) require
increased dust opacities due to large, fluffy grains to explain the
$\leq7$~K temperatures observed in the centers of some low-mass
starless cores. Flower, Pineau des For\^ets, \& Walmsley (2005, 2006)
considered the chemical implications of grain growth, finding it
causes an increase in the electron fraction and the ratio of $\rm
H^+/H_3^+$, and can have moderate effects on the observed differential
freeze-out of nitrogen and carbon bearing species.  Vastel et
al. (2006) required grains larger than $\rm 0.3\mu m$ to reproduce the
ortho-$\rm H_2D^+$ observations in the prestellar core L1544. A
systematic molecular line study of our studied IRDC cores to test for
the above chemical effects and any correlation with $M_{\rm mm}/M_{\rm
SMF}$ would be useful.

%The standard deviation for method1/emission is 0.235112, and the stddev
%for method1/method2 is 0.0594611, all in log10.

%It seems a little odd that our method 2 masses have a larger dispersion
%than method 1, but I suppose it's not that big of a difference, and our
%method 1 masses are obviously less accurate...I guess a lot of it can be
%attributed to the uncertainty in the rathborne masses, right?

\section{Conclusions}

We have presented a new method of diffuse mid-IR ($\rm 8 \mu$m)
extinction mapping of infrared dark clouds to probe the initial
conditions for massive star and star cluster formation. The technique
can naturally probe to higher mass surface densities than NIR
extinction mapping using background stars and provides an estimate of
the mass surface density at scales that are only limited by the
resolution of the mid-IR image (e.g. a couple of arcseconds for {\it
Spitzer} IRAC). However, the method involves a spatial interpolation
and smoothing of the diffuse background, which inevitably introduces
additional uncertainties. In particular it is a relative mass surface
density of the cloud compared to the region used to ascertain the
background that is derived. Also, small scale background fluctuations
are not captured and local regions containing mid-IR bright sources
cannot be treated. One of the largest uncertainties for more distant
clouds is the correction for foreground mid-IR emission from hot dust,
and so we have generally restricted the analysis to relatively nearby
clouds.

We have examined three different methods of estimating the diffuse
background behind IRDCs. The Large-scale Median Filter (LMF) method is the simplest
by not requiring information about the cloud size or location, but
then must smooth over relatively large scales. The Small-scale Median
Filter (SMF) achieves higher spatial resolution of the background given
defined cloud boundaries, and we generally regard this method as being
more accurate where such information is available. Finally, for very
thin, filamentary clouds, more detailed background fitting can be
attempted by considering perpendicular strips across the
filament. Comparing these methods on a sample of 10~IRDCs we find
systematic differences due to background fitting at about the 10\%
level. This means that the mid-IR extinction mapping technique becomes
unreliable for $\Sigma\lesssim 0.013\:{\rm g\:cm^{-2}}$ (given our
adopted fiducial 8~$\rm \mu$m [{\it Spitzer} IRAC Band 4 and diffuse
Galactic background spectrum weighted] opacity of $7.5\:{\rm
cm^{2}\:g^{-1}}$).

We have then used this method to measure the mass surface densities of
the 10 IRDCs. The probability distribution functions of $\Sigma$ for
these clouds show a range of shapes, that is qualitatively consistent
with that expected in numerical simulations of turbulence, though the
relatively narrow distributions of the observed clouds indicate that
the Mach numbers are probably $\lesssim 5$ and/or that magnetic fields
are dynamically important. A more quantitative comparison of the
observed clouds with numerical models, and a search for evolutionary
trends associated with the development of self-gravitating structures
is planned for a future study. 

Given IRDC kinematic distances, we then estimate masses of the clouds
and, more accurately, of their dense cores. Comparing to mass
estimates from mm dust continuum emission (Rathborne et al. 2006) we
find very good agreement: a $\lesssim 10\%$ systematic offset for the
population of 43 studied cores, and a dispersion of less than a factor
of two in the distribution.

Finally, we examine the ratio of masses estimated by mm dust emission
to those found from mid-IR dust extinction as a function of core
density, tentatively finding a trend which can be explained as being
due to opacity changes due to ice mantle formation and grain growth or
by a temperature decrease of about 5~K in the densest cores. Future
studies of the mid-IR extinction law in these cores and molecular line
studies of their chemistry can help to test the reality of this
result and distinguish between these explanations.

\acknowledgements We thank Stella Offner, Richard Klein, Christopher
McKee, Fumitaka Nakamura and Zhi-Yun Li for providing us with mass
surface density data from their numerical simulations. We thank Cara
Battersby, Robert Benjamin, Paola Caselli, Bruce Draine, Eric Ford,
James Jackson, Mark Krumholz, \& Jill Rathborne for helpful
discussions. JCT acknowledges support from NSF CAREER grant
AST-0645412.

%\begin{deluxetable}{lccccccccc}
%emulate
\newpage
\begin{longtable*}{lccccccccc}
\tabletypesize{\footnotesize}
\tablecolumns{9}
\tablewidth{0pt}
\tablecaption{IRDC Cores}
\tablehead{\colhead{Core Name\tablenotemark{a}}                                        &
           \multicolumn{2}{c}{Galactic Coordinates}                                        &
           \colhead{Diameter\tablenotemark{b}}                                               &
           \colhead{$M_{\rm mm}$\tablenotemark{c}}                                &
	   \colhead{$\bar{\Sigma}_{\rm LMF}$}                                      &
           \colhead{$\bar{\Sigma}_{\rm SMF}$}                                      &
	   \colhead{$M_{\rm LMF}$}                                      &
           \colhead{$M_{\rm SMF}$}          &
	   \colhead{$\bar{n}_{\rm H,SMF}$}                           \\
           \colhead{}                                                                      &
           \colhead{$l$}                                                                   &
	   \colhead{$b$}                                                                   &
           \colhead{(pc)}                                                             &
           \colhead{($\sm$)}                                                                 &
           \colhead{($\rm g\:cm^{-2}$)}                                                                 &
           \colhead{($\rm g\:cm^{-2}$)}                                                                 &
           \colhead{($\sm$)}                                                                 &
           \colhead{($\sm$)}                &
	   \colhead{($\rm 10^5 cm^{-3}$)}                                                   }
\startdata

A1 (MM4)  & 18.790  & $-$0.286       &  0.721  & 219 &   0.0829 & 0.0813   & 163 &  157  &     0.232                                            \\
A2 (MM6)  & 18.799  & $-$0.294       &  0.814  &   171 &       0.0805 & 0.0710  & 211 &  188  &  0.193                                             \\
A3 (MM5)  & 18.806  & $-$0.303       &  0.698  &   141 &       0.0805    & 0.0749   & 156 &  142  &   0.230                                               \\
                                                                                               
A4 (MM3)(e)  & 18.701  & $-$0.227    &  0.582  &   194  & - & - & -      &  -   &  -                                                   \\
A5 (MM1)(e)  & 18.735  & $-$0.226    &  0.535  &   736  & - & - &  -      &  -  &  -                                                    \\
A6 (MM2)(e)  & 18.833  & $-$0.299    &  0.465  &   201  & -      & - & - &  -   &  -                                                  \\

\hline
B1 (MM2)  & 19.288  &       0.0824  & 0.431   &  94.5 &         0.0656    &        0.0631   & 57.3  &  51.3  &  0.354               \\
B2 (MM1)  & 19.311  &       0.0675  & 0.337   &  93.7 &         0.0774    &        0.0687   & 32.2  &  25.7  &  0.370                                              \\

\hline
C1 (MM9)    & 28.324  &       0.0677    &  0.824  &  330 &      0.165   &    0.194   & 401  &  539    &   0.532                                            \\    
C2 (MM4)    & 28.345  &       0.0597    &  0.509  &  274 &      0.234   &    0.304   & 239  &  303    &   1.26                                            \\    
C3 (MM11)   & 28.352  &       0.1008    &  0.921  &  371 &      0.199   &    0.235   & 243  &  279    &   0.197                                            \\ 
C4 (MM6)    & 28.355  &       0.0726    &  0.485  &  192 &      0.162   &    0.186   & 204  &  248    &   1.20                                            \\    
C5 (MM14)   & 28.356  &       0.0566    &  0.291  &  30.0  &      0.188   &    0.238   & 66.1  &  71.1  &  1.59                                               \\  
C6 (MM10)   & 28.362  &       0.0532    &  0.824  &  296 &     0.0757    &   0.0857   & 465  &  584    &  0.576                                             \\  
C7 (MM16)   & 28.367  &       0.1203    &  1.07  &  371 &      0.203   &    0.256   & 352  &  406     &   0.183                                          \\   
C8 (MM17)   & 28.388  &       0.0381    &  0.727  &  164 &     0.0789   &   0.0892   & 329  &  387   &   0.556                                            \\        
C9 (MM1)(e) & 28.399  &       0.0812    &  0.630  & 952 &     0.155    &    0.183   & 248 &  291     &   0.643                                           \\
C10 (MM8)(e)  & 28.244  &   0.0127    &  0.751  &  343  & -      &  - & - &  -    &   -                                                 \\
C11 (MM7)(e)  & 28.292  &   0.0065    &  0.582  &  253  & -      & - & - & -      &   -                                               \\
C12 (MM3)(e)  & 28.322 & $-$0.0101    &  0.558  &  400  & -      & - & - & -      &   -                                               \\
C13 (MM5)(e)  & 28.325  &    0.1613   &  0.412  &  147  & -      & - & - & -      &   -                                                \\
C14 (MM12)(e) & 28.328  & $-$0.0388   &  1.02  &  395  & -      & - & - & -       &   -                                              \\
C15 (MM2)(e)  & 28.337  &    0.1170   &  0.533  &  449  & -      & - & - & -      &   -                                               \\
C16 (MM15)(e) & 28.339  &     0.1424  &  0.630  &  111  & -      & - & - & -     &   -                                                \\
C17 (MM18)(e) & 28.417  & $-$0.00726  &  0.824  &  135  & -      & - & - & -      &   -                                               \\
C18 (MM13)(e) & 28.419  &     0.1391  &  1.02  &  385  & -      & - & - & -       &   -                                              \\

\hline
                                                          
D1 (MM5)  & 28.526  & $-$0.2503       & 0.636   &  194 &       0.0873   &    0.0995  & 117 &  126  &  0.271                                              \\ 
D2 (MM7)  & 28.538  & $-$0.2757       & 1.24   &  421 &        0.0799   &   0.0942  & 416 &  441  &   0.127                                             \\ 
D3 (MM3)  & 28.543  & $-$0.2369      & 1.55   &   1690 &       0.0722    &  0.0844  & 672 &  729   &  0.108                                                   \\  
D4 (MM8)  & 28.543  & $-$0.2651       & 0.746   &  154 &       0.0565    &   0.0669  & 119 &  127   &  0.169                                             \\
D5 (MM2)  & 28.559  & $-$0.2274    & 1.55   &  1760 &         0.0817     &     0.0924  & 753 &  813  &  0.121                                                    \\ 
D6 (MM4)  & 28.559  & $-$0.2412      & 1.22   &   663 &     0.0707   &    0.0739 & 324 &  380        &  0.116                                        \\
D7 (MM1)  & 28.565  & $-$0.2350   & 0.912   &   1000 &       0.0603   &      0.0655 & 280 &  318      &  0.232                                               \\
D8 (MM10) & 28.579  & $-$0.2303     & 1.13   &   360 &      0.0708   &     0.0836 & 435 &  447       &  0.171                                        \\
D9 (MM9) & 28.589  & $-$0.2285      & 0.718   &  138 &       0.0829  &      0.0959 & 140 &  160      &  0.239                                        \\
D10 (MM6)  & 28.557  & $-$0.2382       & 0.497   &  98.6 &      -    &     - & - &  -  &     -   \\  

\hline

E1 (MM7)     & 28.644  & 0.1375    & 0.742   &   72.2  &        0.109  &          0.119 & 206 &  230     &  0.311                                           \\  
E2 (MM5)     & 28.650  & 0.1260     & 0.742  &  85.4  &         0.0632   &          0.0774 & 193 &  221  &  0.298                                               \\
E3 (MM2)     & 28.661  & 0.1456    & 0.940   &   327 &         0.0872   &          0.100 & 359 &  398    &  0.264                                                  \\
E4 (MM4)     & 28.717  & 0.1456    & 0.717   &   96.2 &         0.0953  &          0.106 & 140 &  169    & 0.254                                             \\
E5 (MM3)     & 28.580  & 0.1456     & 0.544   &   101 & - & - & -      &  -                         &    -                           \\
E6 (MM6)     & 28.647  & 0.1143    & 0.766   &  93.7  & - & - & -       &  -                         &    -                       \\
E7 (MM1)     & 28.688  & 0.1782    & 0.495   &   119 & - & - & -      &  -                           &    -                         \\

\hline

F1 (MM8)  & 34.422  & 0.24792       &  0.556  &   89.6 &       0.0556    &        0.0497 & 72.8 & 67.8     & 0.217                                           \\ 
F2 (MM7)  & 34.438  & 0.24759       &  0.502  &    72.2 &       0.0573    &        0.0517 & 55.8 & 49.8    & 0.217                                             \\
F3 (MM6)  & 34.448  & 0.25091       &  0.664  &   104 &        0.0625    &        0.0591 & 97.0 & 88.4    & 0.166                                              \\
F4 (MM9)  & 34.454  & 0.25495       &  0.699  &    130 &       0.0583    &       0.0513 & 116 & 89.7       & 0.145                                           \\
F5 (MM5)  & 34.386  & 0.21378       &  0.915  &   550 & - & - & -      &  -                             &   -                   \\
F6 (MM4)  & 34.394  & 0.22156       &  0.431  &   211 &  - & - & -      &  -                            &   -                   \\
F7 (MM2)     & 34.401  & 0.22709    &  0.466  &   1060 &  - & - & -      &  -                            &   -                         \\
F8 (MM1)     & 34.411  & 0.23489    &  0.287  &   985 & - & - & -      &  -                             &   -                        \\
F9 (MM3)     & 34.459  & 0.24867    &  0.431  &    250 & - & - & -      &  -                            &   -                         \\

\hline

G1 (MM3)  & 34.734  &     $-$0.5670  &  0.225  &  11.6  &      0.0606    &        0.0622 & 10.4 &  9.77  &  0.474                                                    \\
G2 (MM2)  & 34.783  &     $-$0.5683  &  0.689  &  156 &        0.0515    &        0.0504 & 112 &  105    &   0.177                                                 \\
G3 (MM4)  & 34.784  &     $-$0.5608  &  0.506  &  58.0  &      0.0667    &        0.0682 & 64.1 &  64.6  &  0.275                                               \\
G4 (MM1)  & 34.712  &     $-$0.5946  &  0.323  &  138 & -  &   - & -      &  -                            &   -                         \\

\hline

H1 (MM9)  & 35.478  & $-$0.3096     &  0.380  &  34.8  &        0.0842   &        0.0869 & 78.7 &  69.9    &   0.705                                           \\ 
H2 (MM4)  & 35.483  & $-$0.2858     &  0.478  &  89.6 &         0.113   &         0.114 & 69.9 &  71.9     &    0.364                                          \\      
H3 (MM5)  & 35.483  & $-$0.2954     &  0.520  &  97.9 &        0.0602    &        0.0621 & 119 &  122      &   0.477                                          \\
H4 (MM8)  & 35.491  & $-$0.2830     &  0.408  &   48.9 &       0.0830   &        0.0854 & 36.0 &  37.4     &    0.304                                         \\ 
H5 (MM6)  & 35.497  & $-$0.2863     &  0.520  &   58.8 &       0.0602    &        0.0629 & 57.8 &  65.3    &   0.257                                            \\   
H6 (MM7)  & 35.522  & $-$0.2724     &  0.548  &   79.6 &        0.133   &         0.138 & 89.8 &  92.0     &   0.308                                         \\ 
H7 (MM2)(e)  & 35.417  & $-$0.2847  &  0.239  &   37.3 & - & - & -      &  -                               &    -                     \\
H8 (MM3)(e)  & 35.452  & $-$0.2951  &  0.337  &   65.5 & - & - & -      &  -                               &    -                     \\
H9 (MM1)(e)  & 35.456  & $-$0.1801  &  0.295  &   63.0 &  - &  - & -      &  -                             &    -                       \\

\hline

I1 (MM1)  & 38.957  & $-$0.4659       &  0.380  &  97.0 &      0.111    &       0.124 & 57.3 &   63.6       &    0.640                                       \\
I2 (MM3)  & 38.971  & $-$0.4588       &  0.157  &   9.12 &     0.135    &       0.142 & 12.8 &   11.6       &    1.65                                       \\
I3 (MM2)(e)  & 38.937  & $-$0.4578    &  0.327  &   60.5 & - & - & -      &  -                              &    -                      \\
I4 (MM4)(e)  & 38.949  & $-$0.4385    &  0.432  &   39.8 & - & - & -      &  -                              &    -                     \\

\hline
\newpage

J1 (MM4)  & 53.128  & 0.0503           &  0.384   & 37.3 &   0.0324    &   0.0219 & 17.1 &   12.2            &    0.118                                  \\   
J2 (MM3)(e)  & 53.092  & 0.1201            &  0.166   & 9.94 & - & - & -      &  -                           &     -                        \\   
J3 (MM5)(e)  & 53.136  & 0.0283            &  0.209   & 10.8 & - & - & -      &  -                           &     -                        \\
J4 (MM1)(e)  & 53.141  & 0.0717            &  0.183   & 103 & - & - & -      &  -                           &     -                        \\
J5 (MM2)(e)  & 53.157  & 0.0672            &  0.288   & 36.5 &  - & - & -     &  -                           &     -                  \\
\enddata                                   
\tablenotetext{a}{Name in parentheses from Rathborne et al. (2006). ``(e)'' indicates their designation of the core as containing $8\:{\rm \mu m}$ emission. We have identified additional cores that contain $8\:{\rm \mu m}$ sources, for which we also do not calculate extinction mass surface densities and masses.}
\tablenotetext{b}{Derived from the angular size of the FWHM of the Gaussian profile fitted to the core by Rathborne et al. (2006).}
\tablenotetext{c}{Millimeter emission mass, from Rathborne et al. (2006).}
\label{tb:cores}
%\end{deluxetable}
%emulate
\end{longtable*}

%$-$$-$$-$$-$$-$$-$$-$$-$$-$$-$$-$$-$$-$$-$$-$$-$$-$$-$$-$$-$$-$$-$$-$$-$$-$$-$$-$$-$$-$$-$$-$$-$$-$$-$$-$$-$$-$$-$$-$$-$$-$$-$$-$$-$$-$$-$$-$$-$$-$$-$$-$$-$$-$$-$$-$$-$$-$$-$$-$$-$$-$$-$$-$$-$$-$$-$$-$$-$$-$$-$$-$$-$$-$$-$

\begin{references}

\reference{} Benjamin, R. A., Churchwell, E., Babler, B. L., Bania, T. M., Clemens, D. P. et al. 2003, \pasp, 115, 953
\reference{} Bergin, E. A., Maret, S., van der Tak, F. F. S., Alves, J., Carmody, S. M., \& Lada, C. J. 2006, \apj, 645, 369
\reference{} Bianchi, S., Goncalves, J., Albrecht, M., Caselli, P., Chini, R., Galli, D., \& Walmsley, M. 2003, \aap, 399, L43
\reference{} Bonnell, I. A., Bate, M. R., \& Zinnecker, H. 1998, \mnras, 298, 93
\reference{} Bonnell, I. A., Bate, M. R., Clarke, C. J., \& Pringle, J. E. 2001, \mnras, 323, 785
\reference{} Bonnell, I. A., Vine, S. G., \& Bate, M. R. 2004, \mnras, 349, 735
\reference{} Carey, S. J., Clark, F. O., Egan, M. P., Price, S. D., Shipman, R. F., \& Kuchar, T. A. 1998, \apj, 508, 721
\reference{} Dalgarno, A., \& Lepp, S. 1984, \apj, 287, L47
\reference{} de Wit, W. J., Testi, L., Palla, F., \& Zinnecker, H. 2005, \aap, 437, 247
\reference{} Draine, B. T. 2003, \araa, 41, 241
\reference{} Egan, M. P., Shipman, R. F., Price, S. D., Carey, S. J., Clark, F. O., Cohen, M. 1998, \apj, 494, L199
\reference{} Elmegreen, B.~G. 2000, \apj, 530, 277
\reference{} Elmegreen, B.~G. 2007, \apj, 668, 1064
\reference{} Flower, D. R., Pineau des F\^orets, G., Walmsley, C. M. 2005, \aap, 436, 933
\reference{} Flower, D. R., Pineau des F\^orets, G., Walmsley, C. M. 2006, \aap, 456, 215
\reference{} Hartmann, L. \& Burkert, A. 2007, \apj, 654, 988
\reference{} Hester, J. J., Desch, S. J., Healy, K. R., \& Leshin, L. A. 2004, Science, 304, 1116
\reference{} Indebetouw, R., et al. 2005, \apj, 619, 931
\reference{} Jackson, J. M. et al. 2006, \apjs, 163, 145
\reference{} Kennicutt, R. C., 1998, \apj, 498, 541
\reference{} Keto, E. \& Caselli, P. 2008, \apj, 683, 238
\reference{} Lada, C. J., \& Lada, E. A. 2003, \araa, 41, 57
\reference{} Li, A., \& Draine, B. T. 2001, \apj, 554, 778
\reference{} Lutz, D. et al. 1996, \aap, 315, L269
\reference{} McKee, C. F., \& Tan, J. C. 2003, \apj, 585, 850
\reference{} McKee, C. F., \& Williams, J. P. 1997, \apj, 476, 144
\reference{} Nakamura, F. \& Li, Z.-Y. 2007, \apj, 662, 395
\reference{} Nakano, T., \& Nakamura, T. 1978, PASJ, 30, 681 
\reference{} Offner, S. S. R., Klein, R. I., McKee, C. F. 2008, \apj, 686, 1174
\reference{} Ossenkopf, V., \& Henning, T. 1994, \aap, 291, 943 
\reference{} P\'erault, M. et al. 1996, \aap, 315, L165
\reference{} Pillai, T., Wyrowski, F., Carey, S. J., \& Menten 2006, \aap, 450, 569
\reference{} Pillai, T., Wyrowski, F., Hatchell, J., Gibb, A. G., \& Thompson, M. A. 2007, \aap, 467, 207
\reference{} Ragan, S. E., Bergin, E. A., Plume, R., Gibson, D. L., Wilner, D. J., O'Brien, S., Hails, E. 2006, \apjs, 166, 567
\reference{} Rathborne, J. M., Jackson, J. M., \& Simon, R. 2006, \apj, 641, 389
\reference{} Rom\'an-Z\'u\~niga, Lada, C. J., Muench, A., \& Alves, J. F. 2007, \apj, 664, 357
\reference{} Sakai, T., Sakai, N., Kamegai, K., Hirota, T., Yamaguchi, N., Shiba, S., Yamamoto, S. 2008, \apj, 678, 1049
\reference{} Shu, F. H., Adams, F. C., \& Lizano, S. 1987, \araa, 25, 23
\reference{} Simon, R., Jackson, J. M., Rathborne, J. M., Chambers, E. T. 2006a, \apj, 639, 227
\reference{} Simon, R., Rathborne, J. M., Shah, R. Y., Jackson, J. M., \& Chambers, E. T. 2006b, \apj, 653, 1325
\reference{} Sridharan, T. K., Beuther, H., Saito, M., Wyrowski, F., \& Schilke P. 2005, \apj, 634, L57
\reference{} Tan, J. C., Krumholz, M. R., McKee, C. F. 2006, \apj, 641, L121
\reference{} Vastel, C., Caselli, P., Ceccarelli, C., Phillips, T., Wiedner, M. C., Peng, R., Houde, M., \& Dominik, C. 2006, \apj, 645, 1198
\reference{} Wang, Y., Zhang, Q., Pillai, T., Wyrowski, F., Wu, Y. 2008, \apj, 672, L33
\reference{} Weingartner, J. C., \& Draine, B. T. 2001, \apj, 548, 296
\end{references}
\end{document}